\documentclass[12pt]{article}
\usepackage{amsmath}
\usepackage{amsfonts}
\usepackage{amssymb}
\usepackage{graphicx}
\usepackage{epsfig}
\newcommand{\be}{\begin{equation}}
\newcommand{\ee}{\end{equation}}
\newcommand{\ben}{\begin{equation*}}
\newcommand{\een}{\end{equation*}}
\newcommand{\bea}{\begin{eqnarray}}
\newcommand{\eea}{\end{eqnarray}}
\newcommand{\ar}{\begin{array}}
\newcommand{\arn}{\end{array}}

\newcommand{\vk}{\vec{k}}

\newcommand{\q}{\vec{q}}
\newcommand{\qs}{\vec{q}^{\;2}}
\newcommand{\qp}{\vec{q}^{\;\prime}}

\def\pnot{\mbox{${\not{\hbox{\kern-3.0pt$p$}}}$}}
\def\qnot{\mbox{${\not{\hbox{\kern-2.0pt$q$}}}$}}
\def\enot{\mbox{${\not{\hbox{\kern-2.0pt$e$}}}$}}
\def\knot{\mbox{${\not{\hbox{\kern-2.0pt$k$}}}$}}

\def\fun#1#2{\lower3.6pt\vbox{\baselineskip0pt\lineskip.9pt\ialign
{$\mathsurround=0pt#1\hfil##\hfil$\crcr#2\crcr\sim\crcr}}}

\begin{document}
\begin{minipage}[t]{3cm}
 DESY-12-161\\
\end{minipage}
\vspace*{1.cm}
\begin{center}
\begin{Large}
NLO Corrections to the kernel of the BKP-equations\\[1cm]
\end{Large}
\vspace{0.5cm}
J. Bartels$^a$, V.S.Fadin$^b$, L.N.Lipatov$^{a,c}$, G.P.Vacca$^d$  \\[1cm] 
\end{center}
$^a$ II. Institut f\"{u}r Theoretische Physik, Universit\"{a}t Hamburg,
Luruper Chaussee 149,\\D-22761 Hamburg, Germany\\
$^b$ Budker Institute of Nuclear Physics, 630090 Novosibirsk, Russia, and
Novosibirsk State University, 630090 Novosibirsk, Russia\\     
$^c$ Petersburg Nuclear Physics Institute, Gatchina 188300, St.Peterburg, Russia
\\
$^d$ INFN Sezione di Bologna, via Irnerio 46, I-40126 Bologna, Italy
\vskip15.0pt 
\noindent
{\bf Abstract:}\\
We present results for the NLO kernel of the BKP equations for composite states of three reggeized gluons in the Odderon channel, both in QCD and in 
N=4 SYM. The NLO kernel consists of the NLO BFKL kernel in the color octet 
representation and the connected $3\to3$ kernel, computed in the tree  
approximation.  
\section{Introduction}
The BFKL equation \cite{BFKL} describing the bound states of two reggeized gluons represents 
one of the fundamental equations in QCD. 
The kernel has been computed in NLO accuracy both for the colour singlet $t$-channel 
state in the forward direction (i.e. at $t=0$) \cite{FL98,Kotikov:2000pm}, and for 
arbitrary $t$ and other possible colour configurations in the $t$-channel \cite{Fadin:1998jv,Fadin:2000kx,Fadin:2004zq}. In
the forward case the solutions of the equation  have been
investigated intensively. Most attention has been given to the color
singlet channel, but recently, motivated by the AdS/CFT
correspondence, also the color octet channel has been investigated
\cite{Bartels:2008sc}.

The generalization of the BFKL equation to bound states consisting of 
three and more reggeized gluons, 
the BKP equation \cite{BKP}, has attracted strong interest, because, in the large-$N_c$ 
limit and in leading order of $\alpha_3$, it has been found to be integrable \cite{int,Lipatov:2009nt}. Integrability plays a vital role 
in analyzing the conjectured correspondence between $N=4$ SYM and string 
theory. A particular case of the BKP equations, the C-odd three 
gluon system, is also of some practical interest: 
in leading order it predicts the existence of a (perturbative) Odderon with intercept 
one \cite{Bartels:1999yt}.
The accuracy of the BKP equations, so far, has been 
limited to the leading logarithmic approximation (LO). Clearly it is important to 
find out whether integrability holds beyond the leading order. Also, the 
Odderon intercept at NLO is of interest: if it would remain at unity, 
it would hint at some deeper property of the Odderon system.    

It is the purpose of this paper to compute the kernel for the BKP equations in the Odderon channel in the NLO approximation. In this approximation 
\begin{equation}
K= K(2\to2)+ K(3\to3)\,,  \label{K=K2+K3}
\end{equation}
where $K(2\to2)$ is the sum over NLO $2\to2$  kernels, and $K(3\to3)$ is the connected $3\to3$ 
kernel computed in the Born approximation.
These building blocks of the NLO kernel are illustrated in Fig.1.
\begin{center}
\epsfig{file=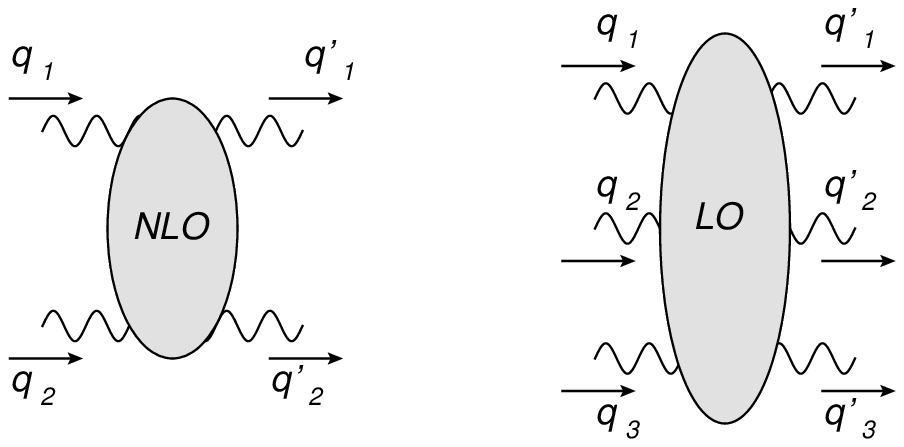,width=8cm,height=4cm}\\
Fig.1: The building blocks of the NLO contributions  to the
BKP equations; $\vec q_i$ and $\vec q^{\;\prime}_i$ are the reggeized gluon momenta transverse 
to the momenta of colliding particles. 
\end{center}  
Since in  the C-odd three
gluon system any pair of two gluons forms  symmetric colour octet subsystems, 
the NLO $2\to2$  kernel  is related to the  NLO BFKL kernel in the symmetric color octet 
representation, though it does not coincide with it. 
We remind that for any colour representation the BFKL kernel can be written as the  
sum of a "virtual" part containing gluon Regge trajectories and  a "real" contribution, 
which results from the $s$-channel intermediate states \cite{BFKL, Fadin:1998fv}. 
In contrast to the BFKL equation which is written for two interacting reggeized gluons and,  
therefore, contains two trajectories and the real contribution, the BKP equation for the Odderon describes the interaction of three reggeized gluons. Thus it contains three trajectories 
and three real parts corresponding to the $s$-channel exchanges between various pairs 
of the reggeized gluons. Writing the $2\to2$ kernel as the sum of the coupled kernels 
\begin{equation}
K(2\to2)=K_{12}+K_{23}+K_{13}\,, \label{K2=K12+K23+K13}
\end{equation}
we obtain that the kernels  $K_{ij}$ have the same real parts as the BFKL kernels for the reggeized 
gluons $i$ and $j$ in the symmetric colour octet representation, but their virtual parts 
are two times smaller. Note that just this property provides 
the infrared stability of  the kernels  $K_{ij}$ as a consequence of the fact that the Odderon 
is a colorless state.

The $3\to3$ kernel is new: in the Born approximation it 
contains only gluons and, hence, it is the same for QCD and for 
supersymmetric Yang-Mills theories. The calculation starts from the effective action 
\cite{EFFACT,Antonov:2004hh} and requires the use of Ward identities of amplitudes 
of reggeized gluons. These Ward identities have been derived recently \cite{BLV}, and they 
play an essential role in the present paper.

This paper consists of two parts. Section 2 contains the NLO BFKL kernel in 
the octet representation, and in section 3 we present the new $3\to3$ vertex.          
Some details of our calculations are collected in an Appendix.  

\section{The $2\to 2$ Kernel}
The $2\to 2$  kernel,  $K(2\to 2)$,   for the three-gluon system is  given by
the sum  (\ref{K2=K12+K23+K13}). Following the discussion after
(\ref{K2=K12+K23+K13}) and  using the representation of the
NLO BFKL kernel given in \cite{Fadin:1998fv} we have, for
the color octet representations,
\begin{equation}
K_{12}^{f,d}= \frac 12 \delta(\vec q_1-\vec q_1^{\;\prime})
\vec q_1^{\;2}\vec q_2^{\;2}\left(\omega(-\vec q_1^{\;2})+\omega(-\vec q_2^{\;2})\right) +
K_r^{f,d}(\vec q_1,\vec q_1^{\;\prime}; \vec q)\,, \label{octkernel}
\end{equation}
where
$\vec q=\vec q_1+\vec q_2=\vec q_1^{\;\prime}+\vec q_2^{\;\prime}$,
$\omega(-\vec q_i^{\;2})$ is the reggeized gluon trajectory,
and $K_r^{f,d}(\vec q_1,\vec q_1^{\;\prime}; \vec q)$ denotes the real part of the
BFKL kernel in the two different octet representation. The superscripts $f$ and $d$ refer
to the antisymmetric ($f$-coupling) and
symmetric ($d$ coupling) adjoint representations, resp. For the Odderon we are
interested in the symmetric octet representation, $K_{12}^d$, but as it will become clear
in the following, it will be easier to start with the antisymmetric case.
It was already mentioned that the kernels $K_{12}$ are infrared stable,
though taken separately the trajectory \cite{Fadin:1996tb} and
the real parts \cite{Fadin:1998jv}, \cite{Fadin:2000kx} are
strongly infrared divergent. It turns out feasible \cite{Fadin:2007de}
to perform the cancelation of the divergencies explicitly and to
write the kernel in the physical two-dimensional transverse momentum space.
To do it one has to use the integral representation for the trajectory.
For QCD such a representation is given in \cite {Fadin:1995xg}.
Keeping in mind that we are interested also in supersymmetric Yang-Mills
theories, we generalize QCD to more general theories with $n_f$
spin-one-half fermions and  $n_s$ scalars in arbitrary representations
of the colour group. For such theories,  using  results of
\cite{Gerasimov:2010zzb}, we obtain in $D=4+2\epsilon$ space time dimensions
\begin{equation}
\omega(-\vec{q}_{i}^{\;2})=\frac{-\bar{g}^{2}\;\vec{q}_{i}^{\;2}}
{\pi^{1+\epsilon}\Gamma(1-\epsilon)}\int\frac{d^{2+2\epsilon}k\;}
{\vec{k}^{\;2}(\vec{k}-\vec{q}_{i})^{2}}
\Biggl(1+\bar{g}^{2}\Biggl[f(\vec
{k},0)+f(0,\vec{k}-\vec{q}_{i})-f(\vec{k},\vec{k}-\vec{q}_{i})\Biggr]\Biggr)~,
\label{omega as double integral}
\end{equation}
where
\begin{equation}
\bar{g}^{2}
=\frac{g^{2}N_{c}\Gamma(1-\epsilon)}{(4\pi)^{2+{\epsilon}}}~.
\label{bar g}
\end{equation}
$\Gamma(x)$ is the Euler gamma-function, $g$ is the bare coupling,
and the function $f$ can be written as
\begin{equation}
f(\vec{k}_{1},\vec{k}_{2})=\frac{\vec{k}_{12}^{\;2}}{\pi^{1+\epsilon}
\Gamma(1-\epsilon)}\int\frac{d^{2+2\epsilon}l}{(\vec{k}
_{1}-\vec{l})^{2}(\vec{k}_{2}-\vec{l})^{2}}\Biggl(\ln\left(  \frac{\vec
{k}_{12}^{\;2}}{\vec{l}^{\;2}}\right)  +a\Biggr)~, \label{f omega}
\end{equation}
with  $\vec{k}_{12}=\vec{k}_{1}-\vec{k}_{2}$. The constant $a$ has the form:
\[
a=-2\psi(1+2\epsilon)-\psi\left(1-\epsilon\right)  +2\psi\left(\epsilon\right)  +\psi(1)+\frac{1}{(1+2\epsilon)}\left(\frac{1}{\epsilon}+\frac{1+\epsilon}{2(3+2\epsilon)}
\right)
\]
\begin{equation}
+\frac{a_f}{N_c}\frac{1+\epsilon}{(1+2\epsilon)(3+2\epsilon)}+\frac{a_s}{N_c}\frac{1}{4(1+2\epsilon)(3+2\epsilon)}~.
\label{a omega}
\end{equation}
Here $\psi(x)=\Gamma^{\prime}(x)/\Gamma(x)$, $a_f=2\kappa_f n_fT_f,
\;\;a_s=2\kappa_s n_s T_s$, $T_f$ and $T_s$  are defined by the relations
\begin{equation}
 \ \mbox{Tr}\left(T_f^aT_f^b\right)=T_f\delta^{ab}, \;\;\; \mbox{Tr}\left(T_s^aT_s^b\right)=T_s\delta^{ab},
\end{equation}
where $T_f^a$ and $T_s^a$ are the colour group generators for
fermions and scalars, respectively, and $\kappa_f$ ($\kappa_s$)
is equal to $1/2$ for Majorana fermions (neutral
scalars) in  self-conjugated representations
and $1$ otherwise. In the case of $n_M$ Majorana fermions
and $n_s$ squarks in the adjoint representation we put $a_f\rightarrow n_M N_c, \;\;a_s\rightarrow n_s N_c$.
For $N$-extended SYM $n_M=N, \;\; n_s=2(N-1)$. In the ${\overline{\mbox{MS}}}$ scheme the bare
coupling is connected with the renormalized coupling, $g_{\mu}$, through the relation
\begin{equation}
g=g_{\mu}\mu^{-\mbox{\normalsize $\epsilon$}}\left[  1+\bar{g}_{\mu}^{2}\frac
{\beta_0}{2N_c\epsilon}\right]  ~,\;\;\bar{g}_\mu^{2}
=\frac{g_\mu^{2}N_{c}\Gamma(1-\epsilon)}{(4\pi)^{2+{\epsilon}}}~, \;\;{\beta_0}{N_c}=\frac{11}{3}{N_c}-\frac{2}{3}a_f-\frac{1}{6}a_s~.
\label{coupling renormalization}
\end{equation}
As it is known,  dimensional regularization violates supersymmetry,
and in supersymmetric theories a modification of dimensional regularization
is used  which is called dimensional reduction. With our accuracy the use of
the dimensional reduction (instead of the dimensional regularization)
is equivalent to the finite charge remormalization
\begin{equation}
\alpha_s(\mu)\rightarrow
\alpha_s(\mu)\left(1-\frac{\alpha_s(\mu)N_c}{12\pi}\right).\label{finite
renormalization}
\end{equation}
The expressions (\ref{f omega}) and (\ref{a omega}) are exact in $\epsilon$.

The real part of the NLO BFKL kernel, $K_r^{f,d}$, contains gluon, fermion, and scalar
contributions. Contrary to the gluon contribution \cite{Fadin:2000kx} to
$K_r$ for which the symmetric and antisymmetric color octet
representations coincide, fermions \cite{Fadin:1998jv} and scalars \cite{Gerasimov:2010zzb}
give different contributions for these representations.
For the fermionic and scalar contributions to $K_r^{d}$ one finds
\be
K^d_{f,s}= K^f_{f,s}+ K^{a}_{f,s},
\ee
such that
\be
K_r^{d}= K_r^{f} +  K^{a}_{s}+  K^{a}_{f}~.
\ee
Here $ K^{a}_{f,s}$ are the so-called abelian parts. They are given by closed
fermion and scalar loops, and
up to coefficients $(g_\mu^4/e^4)b_{f,s}$,
\be
b_{f,s}=\kappa_{f,s}\frac{N_cn_{f,s}}{2(N^2_c-4)(N^2_c-1)}d_{ac_1c_2}d_{ac'_1c'_2}Tr\left(T^{c_1}_{f,s}
T^{c'_2}_{f,s}T^{c_2}_{f,s}T^{c'_1}_{f,s}+T^{c_2}_{f,s}
T^{c'_2}_{f,s}T^{c_1}_{f,s}T^{c'_1}_{f,s}\right)
\ee
they coincide with the Pomeron kernels in usual  QED  \cite{Gribov:1970ik}, \cite{Cheng:1970xm}
and scalar QED. It is important that these pieces are infrared finite,
such that the infrared singular parts of the kernels for
symmetric and antisymmetric adjoint representations are the same.
Moreover, the singular parts of $K_r^f$ and $K_r^d$ are equal to the half of the infrared singular part
of the colour singlet kernel: this makes the procedure of cancellation
of the infrared singularities for all three kernels the same.
Note that the coefficients $b_{f,s}$ are zero
when the colour representations for the particles $f,s$ coincide with that for
the gluons, i.e.  $T^{c}_{ab}=-if_{abc}$ as in $N=4$ SUSY.
This explains why the gluon contribution to $K_r^{f,d}$ is the same for
both adjoint representations. We finally note that at large $N_c$ the nonplanar
contributions  do not contribute.

The contributions   $K^f_{f}$  \cite{Fadin:1998jv} and $K^f_{s}$  \cite{Gerasimov:2010zzb} are known at arbitrary $D$:
\[
K^f_{f}(\vec q_1,\vec q_1^{\;\prime},\vec
q)
=\frac{2\bar{g}^4 a_f}{N_c\pi^{1+\epsilon} \Gamma(1-\epsilon)}\frac{\Gamma^2(2+\epsilon)}{\epsilon \Gamma(4+2\epsilon)}
 \Bigl\{
2\Bigl(\frac{\vec{q_1}{}^2\vec{q_2}'{}^2+
\vec{q_1}'{}^2\vec{q_2}{}^2}{\vec{k}{}^2}-\vec{q}{}^2\Bigr)(\vec{k}^{\;2})^\epsilon+
\]
\[
+\vec q^{\;2}\Bigl(2(\vec{k}^{\;2})^\epsilon+2(\vec{q}^{\;2})^\epsilon- (\vec{q}_1^{\;2})^\epsilon -
(\vec{q_2}^{\;2})^\epsilon-(\vec{q}_1^{\;\prime\;2})^\epsilon - (\vec{q}_2^{\;\prime\;2})^\epsilon \Bigr)
\]
\begin{equation}
- \left(\frac{{\vec{q_1}}{}^2{\vec{q_2}}'{}^2
-{\vec{q_2}}{}^2{\vec{q_1}}'{}^2 }{{\vec{k}}^2}\right)
\left((\vec{q}_1^{\;2})^\epsilon - (\vec{q}_2^{\;2})^\epsilon-(\vec{q}_1^{\;\prime\;2})^\epsilon+
(\vec{q}_2^{\;\prime\;2})^\epsilon\right)\Bigr\},  \label{K F }
\end{equation}
where the total transverse momenta in the $s$ and $t$ channel are given by
\be
\vec{k}=\vec{q}_1-\vec{q}_1^{\;\prime} =\vec{q}_2^{\;\prime}-\vec{q}_2 \label{kdefinition}
\ee
and
\be
\vec{q}=\vec{q}_1+\vec{q}_ 2=\vec{q}_1^{\;\prime}+\vec{q}_2^{\;\prime}, \label{qdefinition}
\ee
resp.,  and
\begin{equation}
K^f_{s}(\vec q_1,\vec q_1^{\;\prime},\vec
q)=\frac{a_s}{4(1+\epsilon)a_f}
K_{f}(\vec q_1,\vec q_1^{\;\prime},\vec
q).
\end{equation}

Turning now to $K_r^f$, in principle it is possible \cite{Fadin:2000kx} to
write the gluon contribution of the real kernel also
at arbitrary $D$. But the result looks too cumbersome.
With the accuracy of keeping, after integration over  $\vec{k}$,
all terms nonvanishing at $\epsilon\rightarrow 0$,
one obtains for the total real kernel in the antisymmetric octet representation, $K_r^f$,
expressed in terms of the renormalized coupling
\[
K^f_r(\vec q_1,\vec q_1^{\;\prime},\vec
q)=\frac{\bar{g}_\mu^{2}\mu^{-2\epsilon}}{\pi^{1+\epsilon}
\Gamma(1-\epsilon)}  \left(
\frac{\vec{q}_{1}^{\;2}\vec{q}_{2}^{\;\prime\;2}+\vec{q}_{1}^{\;\prime\;2}
\vec{q}_{2}^{\;2}}{\vec{k}^{\;2}}-\vec{q}^{\;2}\right)
\]
\[
\times\Biggl(1+\bar{g}_\mu^{2}\left[\frac{\beta_0}{N_c\epsilon
} +\left(\frac{\vec{k}^{\;2}}{\mu^2}\right)^{\epsilon}\left(  -\frac{\beta_0}{N_c\epsilon
  }+\frac{67}{9}-2\zeta(2)- \frac{10}{9}\frac{a_f}{N_c}-\frac{4}{9}\frac{a_s}{N_c} \right.\right.
\]
\begin{equation}
\left.\left.
+\epsilon\left(  -\frac{404}{27}+14\zeta(3)+\frac
{\beta_0}{N_c}\zeta(2)+\frac{56}{27}\frac{a_f}{N_c}+\frac{26}{27}
\frac{a_s}{N_c}\right) \right) \right]\Biggr) +\frac{\bar{g}_\mu^{4}}{2\pi}R(\vec q_1,\vec q_1^{\;\prime},\vec
q),
\label{K^f}
\end{equation}
where $\vec k$ is given in (\ref{kdefinition}). The  function $R(\vec q_1,\vec q_1^{\;\prime},\vec q)$ is infrared safe
and can be evaluated at $D=4$:
\[
R(\vec q_1,\vec q_1^{\;\prime},\vec q) =
\Biggl[\vec{q}^{\,2}\left(  \frac{\beta_0}{N_c}\ln\left(
\frac{\vec{q}_{1}^{\;2}\vec{q}_{1}^{\;\prime\;2}}{\vec{q}^{\;2}\vec{k}^{\;2}
}\right)  +\frac{1}{2}\ln\left(  \frac{\vec{q}_{1}^{\;2}}{\vec{q}^{\;2}
}\right)  \ln\left(  \frac{\vec{q}_{2}^{\;2}}{\vec{q}^{\;2}}\right)  +\frac
{1}{2}\ln\left(  \frac{\vec{q}_{1}^{\;\prime\;2}}{\vec{q}^{\;2}}\right)
\ln\left(  \frac{\vec{q}_{2}^{\;\prime2}}{\vec{q}^{\;2}}\right)  \right.
\]
\[
\left.  +\frac{1}{2}\ln^{2}\left(  \frac{\vec{q}_{1}^{\;2}}{\vec{q}
_{1}^{\;\prime\;2}}\right)  \right)  -\frac{\vec{q}_{1}^{\;2}\vec{q}
_{2}^{\;\prime\;2}+\vec{q}_{2}^{\;2}\vec{q}_{1}^{\;\prime\;2}}{\vec{k}^{\;2}
}\ln^{2}\left(  \frac{\vec{q}_{1}^{\;2}}{\vec{q}_{1}^{\;\prime\;2}}\right)
+\frac{\vec{q}_{1}^{\;2}\vec{q}_{2}^{\;\prime\;2}-\vec{q}_{2}^{\;2}\vec{q}
_{1}^{\;\prime\;2}}{\vec{k}^{\;2}}\ln\left(  \frac{\vec{q}_{1}^{\;2}}{\vec
{q}_{1}^{\;\prime\;2}}\right)
\]
\[
\times \left(  \frac{\beta_0}{N_c}-\frac{1}{2}\ln\left(
\frac{\vec{q}_{1}^{\;2}\vec{q}_{1}^{\;\prime\;2}}{\vec{k}^{\;4}}\right)
\right)
+4\frac{(\vec k\times \vec q_1)}{\vec{k}^{\;2}}\left(\vec{k}^{\;2}(\vec q_1\times \vec q_2)-\vec{q_1}^{\;2}(\vec k \times \vec q_2)-\vec{q_2}^{\;2}(\vec k \times \vec q_1)\right)I_{\vec q_1,-\vec k}
\]
\begin{equation}
+\left(  \vec{q}_{1}\leftrightarrow -\vec{q}_{2},\;\;\vec{q}_{1}^{\;\prime
}\leftrightarrow - \vec{q}_{2}^{\;\prime}\right)  ~.  \label{real octet with g mu}
\end{equation}
Here $(\vec a\times\vec b)= a_x b_y -a_y b_x$ and
\begin{equation}
I_{\vec p, \vec q}=
\int_{0}^{1}\frac{dx}{(\vec p +x\vec q)^{2}}\ln\left(\frac{\vec p^{\;2}}
{x^2\vec q^{\;2}}\right)~ \label{I p q 1}
\end{equation}
can be expressed in terms of the di-logarithmic function and has the following symmetry properties:
\begin{equation}
I_{\vec p,\vec q}=I_{-\vec p,-\vec q}=I_{\vec q, \vec p}=I_{\vec p,-\vec p
-\vec q}~.
\label{I definition}
\end{equation}
This symmetry is evident from the  representation
\begin{equation}
I_{\vec p,\vec q}=\int_{0}^{1}\int_{0}^{1}\int_{0}^{1}\frac{dx_{1}dx_{2}dx_{3}
\delta(1-x_{1}-x_{2}-x_{3})}{(\vec p^{\;2}x_{1}+\vec q^{\;2}x_{2}+(\vec p
+\vec q)^{2}x_{3})(x_{1}x_{2}+x_{1}x_{3}+x_{2}x_{3})}~.\label{I symmetric}
\end{equation}
Other useful representations are
\[
I_{\vec p,\vec q}=\int_{0}^{1}
\frac{dx}{a(1-x)+bx-c x(1-x)}\ln\left(  \frac{a(1-x)+bx}{cx(1-x)}\right)
\]
\begin{equation}
=\int_{0}^{1}dx\int_{0}^{1}{dz}\;\frac{1}{cx(1-x)z+(b(1-x)+ax)(1-z)}~,
\label{integral I}
\end{equation}
where $a=\vec p^{\;2}, b=\vec q^{\;2}, c = (\vec p+\vec q)^{2} $.

Keeping only the terms of zero order in $\epsilon$ in the trajectory (\ref{omega as double integral}) and passing to the renormalized  coupling, one obtains
\begin{equation}
\omega(-\vec{q}_{i}^{\;2})=-\frac{\bar{g}_\mu^{2}\;\vec{q}_{i}^{\;2}}
{\pi^{1+\epsilon}\Gamma(1-\epsilon)}\int\frac{d^{2+2\epsilon}k\,\mu^{-2\epsilon}}{\vec{k}^{\;2}(\vec{k}-\vec{q}_{i})^{2}}\left(  1+\bar{g}_\mu^{2}f_\mu^{\omega}(\vec{k},\vec{k}-\vec{q}_{i})\right)  ~,
\label{omega with g mu as integral}
\end{equation}
where
\[
f_\mu^{\omega}(\vec{k}_{1},\vec{k}_{2})=\frac{\beta_0}{N_c\epsilon}+\left[
\frac{\beta_0}{N_c\epsilon}-\frac{67}{9}
+2\zeta(2)+\frac{10}{9}\frac{a_f}{N_c} \right.
\]
\[
\left.
+\frac{4}{9}\frac{a_s}{N_c} +\epsilon\left(  \frac{404}{27}-\frac{\beta_0}{N_c}\zeta(2)-6\zeta(3)-\frac{56}{27}\frac{a_f}{N_c} -\frac{26}{27}\frac{a_s}{N_c}\right)
\right]
\]
\begin{equation}
\times
 \left[  \left(  \frac{\vec{k}_{12}^{\;2}}{\mu^2}\right)
^{\epsilon}-\left(  \frac{\vec{k}_{1}^{\;2}}{\mu^2}\right)  ^{\epsilon
}-\left( \frac{\vec{k}_{2}^{\;2}}{\mu^2}\right)^{\epsilon}\right]
-\ln\left(  \frac{\vec{k}_{12}^{\;2}}{\vec{k}_{1}^{\;2}}\right)  \ln\left(
\frac{\vec{k}_{12}^{\;2}}{\vec{k}_{2}^{\;2}}\right)  ~.
\label{integrand for omega with g mu}
\end{equation}
This representation allows to calculate $\omega(-\vec{q}^{\;2})$  explicitly:
\[
\omega(-\vec{q}^{\;2}) = -\bar g^2_\mu\left(\frac{2}{\epsilon} + 2\ln\left(
\frac{\vec{q}^{\;2}}{\mu^2}\right)\right)-\bar g^4_\mu\left[\frac{\beta_0}{N_c}
\left(\frac{1}{\epsilon^2}-\ln^2\left(
\frac{\vec{q}^{\;2}}{\mu^2}\right)\right)+ \left(\frac{1}{\epsilon}+2\ln\left(\frac{\vec{q}^{\;2}}{\mu^2}\right)\right)\right.
\]
\begin{equation}
\left.\left. \times (\frac{67}9-2\zeta(2)-\frac{10}9\frac{a_f}{N_c}-\frac{4}9\frac{a_s}{N_c}\right)
-\frac{404}{27}+2\zeta(3)+\frac{56}{27}\frac{a_f}{N_c}+\frac{26}{27}\frac{a_s}{N_c}\right]. \label{omega expanded}
\end{equation}

In order to cancel the infrared singularities and
to find  the kernel (\ref{octkernel}) at $D=4$ let us introduce
the cut-off $\lambda\rightarrow0$ which should be  taken to zero after pushing  $\epsilon\rightarrow0$.
In the integral representation of the trajectory
(\ref{omega with g mu as integral}) we divide the integration region
into three domains. In two of them either $\vec
{k}^{\;2}\leq\lambda^{2}$, or $(\vec{k}-\vec{q}_{i})^{2}\leq\lambda^{2}$, and
in the third one both $\vec{k}^{\;2}>\lambda^{2}$ and $(\vec{k}-\vec{q}
_{i})^{2}>\lambda^{2}$. Then in the third domain we can take
$\epsilon=0$ in (\ref{integrand for omega with g mu}) and
put $f^{\omega}_\mu(\vec{k}
_{1},\vec{k}_{2})=f_{\omega}^{(0)}(\vec{k}_{1},\vec{k}_{2})$, where
\begin{equation}
f_{\omega}^{(0)}(\vec{k}_{1},\vec{k}_{2})=\frac{67}{9}-2\zeta(2)-\frac{10}{9}\frac{a_f}{N_c}
-\frac{4}{9}\frac{a_s}{N_c}
-\frac{\beta_0}{N_c}\ln\left(  \frac{\vec{k}_{1}^{\;2}\vec{k}_{2}^{\;2}}{\mu^{2}\vec{k}
_{12}^{\;2}}\right)  -\ln\left(  \frac{\vec{k}_{12}^{\;2}}{\vec{k}_{1}^{\;2}
}\right)  \ln\left(  \frac{\vec{k}_{12}^{\;2}}{\vec{k}_{2}^{\;2}}\right)  ~.
\label{integrand for omega at large k}
\end{equation}
In the first domain we have
\[
f^{\omega}_\mu(\vec{k},\vec{k}-\vec{q}_{i})=\frac{\beta_0}{N_c\epsilon}-\left(
\frac{\vec{k}^{\;2}}{\mu^{2}}\right)  ^{\epsilon}\Biggl[\frac{\beta_0}{N_c\epsilon}-\frac{67}{9}+2\zeta(2)+\frac{10}{9}
\frac{a_f}{N_c}+\frac{4}{9}\frac{a_s}{N_c}
\]
\begin{equation}
+ \epsilon\left(  \frac{404}{27}-\frac{\beta_0}{N_c}\zeta(2)-6\zeta(3)-\frac{56}{27}\frac{a_f}{N_c} -\frac{26}{27}\frac{a_s}{N_c}\right) \Biggr]~, \label{integrand for omega at small k}
\end{equation}
and in the second one we have the same expression with the substitution
$\vec{k}^{\;2}\rightarrow(\vec{k}-\vec{q}_{i})^{2}$.

Writing the real part of the antisymmetric color octet kernel as
\begin{equation}
K_r^f(\vec q_1, \vec q_1^{\;\prime},\vec q)=K_r^f(\vec q_1, \vec q_1^{\;\prime},\vec q)\theta(\lambda^{2}-\vec{k}^{\;2})+K_r^f(\vec q_1, \vec q_1^{\;\prime},\vec q)\theta(\vec{k}^{\;2}-\lambda^{2})~,
\label{decomposition of K 8r}
\end{equation}
and comparing (\ref{K^f}) with (\ref{integrand for omega at small k}),
we see that in the  kernel $K_{12}^f$
(\ref{octkernel}) the first term in the rhs of
(\ref{decomposition of K 8r}) cancels almost completely against  the contributions of
the regions $\vec{k}^{\;2}\leq\lambda^{2}$ and $(\vec{k}-\vec{q}_{i})^{2}
\leq\lambda^{2}$ in the trajectories $\omega(-\vec{q}_{i}^{\;2})$. The only
piece which remains uncanceled in each of the trajectories for $\epsilon
\rightarrow0$ is
\begin{equation}
\frac{\bar{g}_{\mu}^{4}\;}{\pi^{1+\epsilon}\Gamma(1-\epsilon)}\int
\frac{d^{2+2\epsilon}k\;\mu^{-2\epsilon}}{\vec{k}^{\;2}}8\epsilon
\zeta(3)\left(  \frac{\vec{k}^{\;2}}{\mu^{2}}\right)  ^{\epsilon}
\theta(\lambda^{2}-\vec{k}^{\;2})=\frac{\alpha_{s}^{2}(\mu)N_{c}^{2}}{4\pi
^{2}}\zeta(3). \label{uncancelled piece of omega}
\end{equation}
On account of this cancellation and using the equality
\begin{equation}
\int\frac{d^{2}k\;}{4\pi}\frac{\vec{q}^{\;2}}{\vec{k}^{\;2}(\vec{k}-\vec
{q})^{2}}\ln\left(  \frac{\vec{k}^{\;2}}{\vec{q}^{\;2}}\right)  \ln\left(
\frac{(\vec{k}-\vec{q})^{2}}{\vec{q}^{\;2}}\right)  =\zeta(3)~,
\label{representation for zeta}
\end{equation}
we can obtain
\[
K^f_{12}=-\delta(\vec{q}_{1}-\vec{q}_{1}^{\;\prime})
\vec{q}_1^{\;2}\vec{q}_2^{\;2}
\frac{\alpha_{s}(\mu)N_{c}}{8\pi^{2}}\Biggl( \int d^{2}k\;\Biggl[ \frac{2}{\vec{k}^{\;2}}+2\frac{\vec{k}
(\vec{q}_{1}-\vec{k})}{\vec{k}^{\;2}(\vec{q}_{1}-\vec{k})^2}
\]
\[
+\frac{\alpha
_{s}(\mu)N_{c}}{\pi}\left(  V(\vec{k})+V(\vec{k},\vec{k}-\vec{q}_{1})\right)\Biggr]-3\alpha_{s}(\mu)N_{c}\zeta(3)
 \Biggr)
\]
\[
+\frac{\alpha_{s}(\mu)N_{c}}{8\pi^{2}}\left\{  \left(  \frac{\vec{q}_{1}^{\;2}\vec
{q}_{2}^{\;\prime\;2}+\vec{q}_{1}^{\;\prime\;2}\vec{q}_{2}^{\;2}}{\vec
{k}^{\;2}}-\vec{q}^{\;2}\right)  \Biggl(1+\frac{\alpha_{s}(\mu)N_{c}}{4\pi
}\Biggl[-\frac{\beta_0}{N_c}\ln\left(  \frac{\vec{k}^{\;2}}{\mu^{2}}\right)
+\frac{67}{9}\right.
\]
\begin{equation}
\left.-2\zeta(2)-\frac{10}{9}\frac{a_f}{N_c}-\frac{4}{9}\frac{a_s}{N_c}\Biggr]
\Biggr)\!\right\}+\frac{\alpha^2_{s}(\mu)N^2_{c}}{4\pi}R(\vec q_1,\vec q_1^{\;\prime},\vec
q)
+\left(  \vec{q}_{1}\leftrightarrow -\vec{q}_{2},\;\;\vec{q}_{1}^{\;\prime}
\leftrightarrow -\vec{q}_{2}^{\;\prime}\right)  ~, \label{ns kernel at D=4}
\end{equation}
where
\begin{equation}
V(\vec{k})=\frac{1}{2\vec{k}^{\;2}}\left(  \frac{67}{9}-2\zeta(2)-\frac{10}{9}\frac{a_f}{N_c}-\frac{4}{9}\frac{a_s}{N_c}-\frac{\beta_0}{N_c}\ln\left(  \frac{\vec{k}^{\;2}}{\mu^{2}}\right)  \right)  ~,
\label{function V1}
\end{equation}
\[
V(\vec{k},\vec{q})=\frac{\vec{k}\vec{q}}{2\vec{k}^{\;2}\vec{q}^{\;2}}\left(
 \frac{\beta_0}{N_c}\ln\left(  \frac{\vec{k}^{\;2}\vec{q}^{\;2}}{\mu^{2}(\vec{k}
-\vec{q})^{2}}\right)  -\frac{67}{9}+2\zeta(2)+\frac{10}{9}\frac{a_f}{N_c}+\frac{4}{9}\frac{a_s}{N_c} \right)
\]
\begin{equation}
 -\frac{\beta_0}{4N_c\vec
{k}^{\;2}}\ln\left(  \frac{\vec{q}^{\;2}}{(\vec{k}-\vec{q})^{2}}\right)
-\frac{\beta_0}{4N_c\vec{q}^{\;2}}\ln\left(  \frac{\vec{k}^{\;2}}{(\vec{k}-\vec
{q})^{2}}\right)  ~. \label{function V2}
\end{equation}
Of course, the infrared singularities in (\ref{ns kernel at D=4}) must be regularized
either by limitations on the integration regions as discussed above or in an
equivalent way.

Thus we have obtained  the antisymmetric octet kernel, $K_{12}^f$, at $D=4$. For the Odderon which belongs to the symmetric representation ($d$-coupling),
the kernel $K_{12}^d$ differs from $K^f_{12}$ only by the abelian parts of the
fermion and  scalar contributions,
\be
K_{12}^d= K^f_{12}+  K^{a}_{f}+  K^{a}_{s},
\ee
which have no infrared singularities and can directly be taken at $D=4$.
However, the corresponding expressions are rather complicated. The fact that for the
antisymmetric adjoint representation the kernel was relatively simple
is connected with the gluon reggeization and with the absence of the
contribution of non-planar diagrams in this kernel.
In contrast to this, the abelian parts contain such
nonplanar contributions (crossed fermion and scalar loops).

From the results of \cite{Fadin:1998jv} and  \cite{Fadin:2007ee} it follows:
\[
K^{a}_{f}(\vec q_1, \vec q_1^{\;\prime},\vec q)=
\frac{\alpha_s^2(\mu)b_f}{4\pi^3}\int_0^1 dx\int
\frac{d^2l_1}{\pi}\left[2 x(1-x) \frac{2(
\vec{q}_{1}\vec{l}_{1})-\vec{q}_{1}^{\:2}}{\sigma_{11}}
\left( \frac{2(\vec{q}_{2}\vec{l}_{1})+\vec{q}
_{2}^{\:2}}{\sigma_{12}}+\frac{2(\vec{q}_{2}
\vec{l}_{2})+\vec{q}_{2}^{\:2}}{\sigma_{22}}\right)\right.
\]
\[
 + \frac{x
\vec{q}^{\:2}(\vec{q}_{1}^{\:2}-2(\vec{q}_{1}\vec{l}_{1}))}{\sigma
_{11}\sigma_{12}} +\frac{1}{\sigma_{11}\sigma_{22}}\biggl(
-4(\vec{q}_{1}\vec{l}_{1})(\vec{q}_{2}\vec{q}_{2}^{\:\prime})
-2(\vec{l}_{1}\vec{k})\vec{q}_{1}^{\:2}+2 (\vec{q}_{1}\vec{k})(\vec{q}_{2}\vec{k})
+{\vec{q}_{1}^{\:2}\vec{q}_{2}^{\:2}}
\]
\begin{equation}
\left.+\frac{{\vec{k}}_{{}}^{2}(\vec{q}_1-\vec{q}_2)^{2}}{ 2}+x\vec{q}^{\:2}(2(
\vec{q}_{1}\vec{l}_{1})-\vec{q}_{1}^{\:2})\biggr)\right]
+(\vec{q}_{1}\leftrightarrow \vec{q}_{2},\;  \vec{q}_{1}^{\:\prime}\leftrightarrow \vec{q}_{2}^{\:\prime})\;,
\label{F in p space} 
\end{equation}
where 
\be
\vec{l}_1+\vec{l}_2=\vk=\q_1-\qp_1=\qp_2-\q_2,
\ee
\[
\sigma_{11}=(\vec{l}_1-x\q_1)^2+x(1-x)\qs_1,
\;\;\;\sigma_{12}=(\vec{l}_1+x\q_2)^2+x(1-x)\qs_2,
\]
\begin{equation}
\;\;\sigma_{22}=(\vec{l}_2+(1-x)\q_2)^2+x(1-x)\qs_2\;. \label{sigma}
\end{equation}
Performing the integration over $\vec{l}_1$ with the help of Feynman parametrization, we
obtain
\[
K^{a}_{f}(\vec q_1, \vec q_1^{\;\prime},\vec q)=
\frac{\alpha_s^2(\mu)b_f}{4\pi^3}\int_0^1 dx\,
\int_0^1dz\Biggl[-2x(1-x)\biggl( 2\vec q_1\vec q_2
\ln\left(\frac{R_{11}}{R_{12}}\right)
\]
\[
+\frac{\left(\vec q_1^{\; 2}-2\vec r_{11}\vec q_1\right)\left(\vec
q_2^{\; 2}+2\vec r_{11}\vec q_2\right)}{R_{11}}+ \frac{\left(\vec
q_1^{\; 2}-2\vec r_{12}\vec q_1\right)\left(\vec
q_2^{\;\prime\,2}-\vec k^{\; 2}-2\vec r_{12}\vec
q_2\right)}{R_{12}}\biggr)
\]
\[
+\frac{x\vec q^{\; 2}\left(\vec q_1^{\; 2}-2\vec r_{11}\vec q_1\right)}{R_{11}}
+\frac{1}{R_{12}}\biggl(
-4(\vec{q}_{2}\vec{q}_{2}^{\:\prime})(\vec{q}_{1}\vec{r}_{12})
-2\vec{q}_{1}^{\:2}(\vec{r}_{12}\vec{k})+ 2(\vec{q}_{1}\vec{k})(\vec{q}_{2}\vec{k})
+{\vec{q}_{1}^{\:2}\vec{q}_{2}^{\:2}}
\]
\begin{equation}
+\frac{{\vec{k}}_{{}}^{2}(\vec{q}_1-\vec{q}_2)^{2}}{ 2}+x\vec{q}^{\:2}(2(
\vec{q}_{1}\vec{r}_{12})-\vec{q}_{1}^{\:2})\biggr)\Biggr]~+(\vec{q}_{1}\leftrightarrow \vec{q}_{2},\;  \vec{q}_{1}^{\:\prime}\leftrightarrow \vec{q}_{2}^{\:\prime})\;,
\end{equation}
where
\[
\vec r_{11}=x\left(z\vec q_1-(1-z)\vec q_2\right)~, \;\;\vec
r_{12}=xz\vec q_1+(1-x)(1-z)\vec q_2 +(1-z)\vec k~,
\]
\[
R_{11}=x(1-x)\left(z\vec q_1^{\; 2}+(1-z)\vec q_2^{\;
2}\right)+x^2z(1-z)\vec q^{\; 2}, \;\;
\]
\begin{equation}
R_{12}=x(1-x)\left(z\vec q_1^{\; 2}+(1-z)\vec q_2^{\;
2}\right)+z(1-z)\left(x\vec q_1^{\;\prime\, 2}+(1-x)\vec
q_2^{\;\prime\, 2} -x(1-x)\vec q^{\; 2}\right)~.
\end{equation}
Further integration leads to a lengthy expression which will not be presented here.

For the scalar contribution we have, using the results of \cite{Gerasimov:2010zzb}:
\[
K^{a}_{s}(\vec q_1, \vec q_1^{\;\prime},\vec q)=
\frac{\alpha_s^2(\mu)b_s}{4\pi^3}
\int_0^1 dx\int\frac{d\vec{l}_1}{\pi}\,\,x(1-x)
\frac{\vec{q_1}^2-2(\vec{l_1}\vec{q_1})}{\sigma_{11}}
\]
\begin{equation}
\times\left(\frac{\vec{q_2}{}^2+2(\vec{l_1}\vec{q_2})}{\sigma_{12}}
+\frac{\vec{q_2}{}^2+2(\vec{l_2}\vec{q_2})}{\sigma_{22}} \right) +(\vec{q}_{1}\leftrightarrow \vec{q}_{2},\;  \vec{q}_{1}^{\:\prime}\leftrightarrow \vec{q}_{2}^{\:\prime})~,
\label{k_a_2}
\end{equation}
and after integration over $\vec{l}_1$
\[
K^{a}_{s}(\vec q_1, \vec q_1^{\;\prime},\vec q)=
\frac{\alpha_s^2(\mu)b_s}{2\pi^3}
\int_0^1 dx\,
x(1-x)\int_0^1dz\Biggl[ 2\vec q_1\vec q_2
\ln\left(\frac{R_{11}}{R_{12}}\right)
\]
\begin{equation}
+\frac{\left(\vec q_1^{\; 2}-2\vec r_{11}\vec q_1\right)\left(\vec
q_2^{\; 2}+2\vec r_{11}\vec q_2\right)}{R_{11}}+ \frac{\left(\vec
q_1^{\; 2}-2\vec r_{12}\vec q_1\right)\left(\vec
q_2^{\;\prime\,2}-\vec k^{\; 2}-2\vec r_{12}\vec
q_2\right)}{R_{12}}\Biggr]~.
\end{equation}
The abelian contributions are drastically simplified if we transform them into the space of transverse coordinates 
and choose the M\"{o}bius representation~\cite{Bartels:2004ef}.
The M\"{o}bius form of these contributions  can be written~\cite{Fadin:2011zz} as
\[
\langle\vec{r}_{1}\vec{r}_{2}|\hat{\mathcal{K}}^a_{M}|\vec{r}
_{1}^{\;\prime}\vec{r}_{2}^{\;\prime}\rangle=\frac{\alpha^2_s} {2\pi^4}
\frac{1}{\vec{r}_{1^{\prime}2^{\prime}}^{\,\,4}}\left[\left(  \frac{\vec
{r}_{11^{\prime}}^{\;2}\,\vec{r}_{22^{\prime}}^{\;2}-2\vec{r}_{12}^{\;2}%
\,\vec{r}_{1^{\prime}2^{\prime}}^{\;2}}{d}\ln\left(  \frac{\vec{r}%
_{12^{\prime}}^{\;2}\,\vec{r}_{21^{\prime}}^{\;2}}{\vec{r}_{11^{\prime}}%
^{\;2}\vec{r}_{22^{\prime}}^{\;2}}\right)  -1\right)  \left({b_{s}}-2b_f\right)\right.
\]%
\begin{equation}
\left.+  \frac{(2b_{s}-3b_{f})}{\vec{r}_{1^{\prime}2^{\prime}}^{\,\,2}}%
\frac{\vec{r}_{12}^{\;2}\,}{d} \ln\left(  \frac{\vec{r}_{12^{\prime}}^{\;2}\,\vec{r}_{21^{\prime}%
}^{\;2}}{\vec{r}_{11^{\prime}}^{\;2}\vec{r}_{22^{\prime}}^{\;2}}\right)\right]~.
\end{equation}
Here  $\vec{r}_{ij^\prime}=\vec{r}_{i}-\vec{r}_{j}^{\;\prime}$.

Finally, the total $2\to 2$ kernel (\ref{K2=K12+K23+K13}) for the Odderon is expressed
in terms of  $K_{ij}^d$. It is worthwhile to note here that the real 
part of the kernel turns to zero at zero transverse momentum of any of the reggeons.

\section{The $3 \to 3$ kernel}
The kinematics of the lowest order $3 \to 3$ transition is illustrated in Fig.2. 
The grey blobs on the right and on the left hand side are impact factors 
for the external projectiles $A$, and $B$,  
the horizontal wavy lines denote reggeized gluons, 
and the black ellipse in the center stands for  
the $3 \to 3 $  vertex we are interested in.  
\begin{center}
\epsfig{file=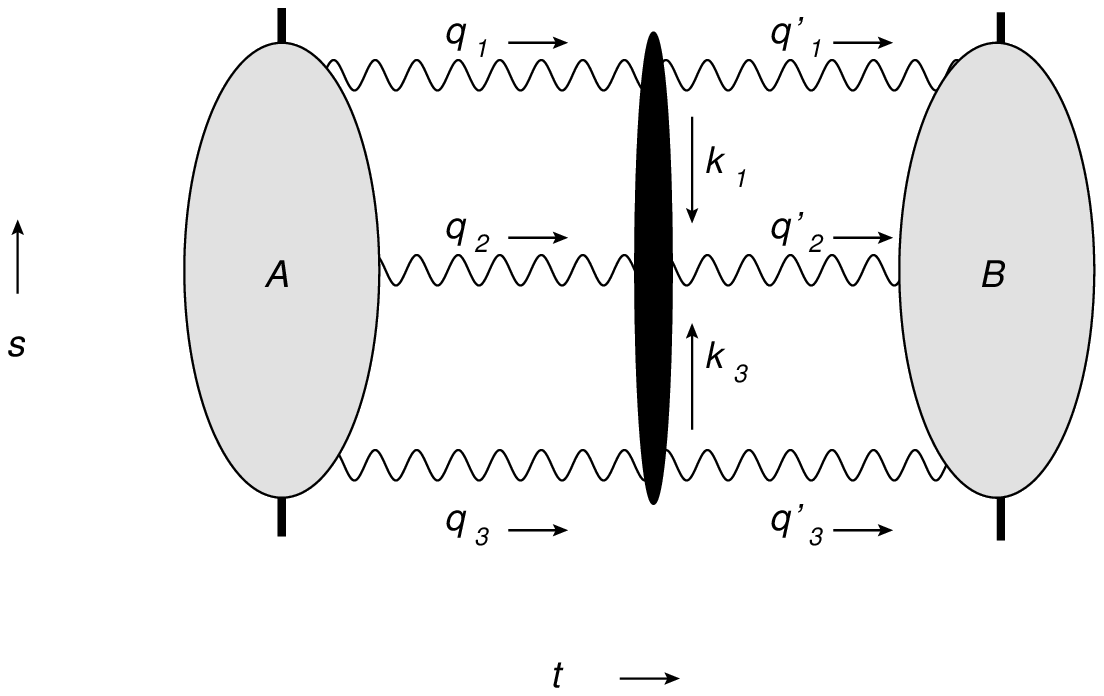,width=8cm,height=6cm}\\
Fig.2: Kinematics of the $3 \to 3$ transition of reggeized gluons. 
\end{center}
In this section the derivation of this connected $3 \to 3$ vertex in lowest order will be described 
in some detail. We make use of the effective action \cite{EFFACT,Antonov:2004hh} and of the Ward identities discussed in \cite{BLV}.

We introduce the Sudakov decomposition of the reggeon momenta
\be 
q_i^{\mu} = \frac{1}{2} (n^-)^{\mu} q_i^+ +   \frac{1}{2} (n^+)^{\mu} q_i^- + q_{i\perp}^{\mu},
\ee
where the lightlike vectors $n^-$ and $n^+$ have the property  
\be
(n^-)^2 = (n^+)^2 =0,\; n^- n^+ =2.
\ee 
The gluon vertices and diagrams we need to include\footnote{We remind that in this paper we restrict 
ourselves to 3-gluon states belonging to the symmetric color representation. 
We therefore do not include diagrams 
which, for example, are antisymmetric in the pair of gluons '1' and '2'.}   
are illustrated in Fig.3
which shows the building blocks derived form the effective action in \cite{Antonov:2004hh}:
(a) the effective RPR-production vertex,  (b) the PPR vertex, (c) the RPRR, and (d) the RRPR, and (e) the RPPR vertex. It is important to emphasize that 
the gluons corresponding to the vertical $s$-channel lines are allowed to be off-shell.
In Fig.4  we illustrate the full $3\to3$ kernel, built from the effective vertices in Fig.3. 
In addition, we have to sum over all permutations of the reggeized gluons: 
in  order to avoid double counting it is enough to add, on the lhs of the kernel, the three 
sequences  (123), (231), and (312) of the lines $q_1$, $q_2$, and $q_3$ (counting from the top to the bottom). On the rhs, we sum over all six permutations of the lines $q'_1$, $q'_2$ and $q'_3$. 
As indicated in Fig.2, in the context of the odderon kernel the $3 \to 3$ vertex will be convoluted with impact factors on both sides. These impact factors depend upon the transverse momenta; the one on the rhs contains the integrals over the longitudinal components ${q'_i}^+$, the one on the lhs is integrated over the ${q_i}^-$. As functions of color and of  the transverse momenta, the impact factors are symmetric under the exchange of gluon lines. In the case of the odderon, they are proportional to the       
symmetric structure constants $d_{abc}$. When contracted with the color tensors 
of the $3 \to 3$ vertex, the nonplanar 3rd term in Fig.3e does not contribute. 
The nonplanar contribution 
also drops out if we insert the $3 \to 3$ vertex into the BKP evolution with overall vacuum  
quantum numbers and take the limit of large $N_c$.  

\begin{center}
\epsfig{file=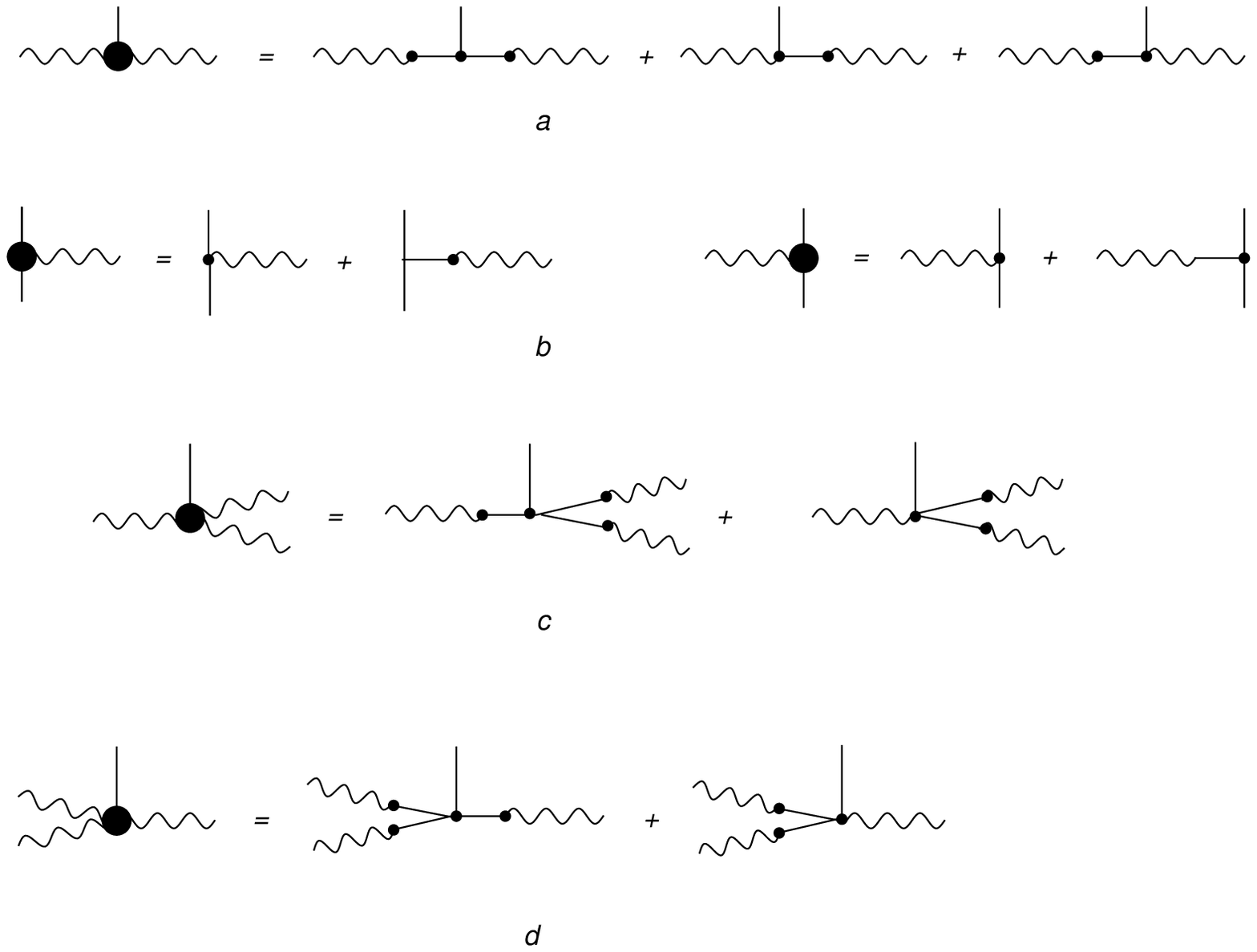,width=12cm,height=8cm}\\
\vspace{-2cm}\epsfig{file=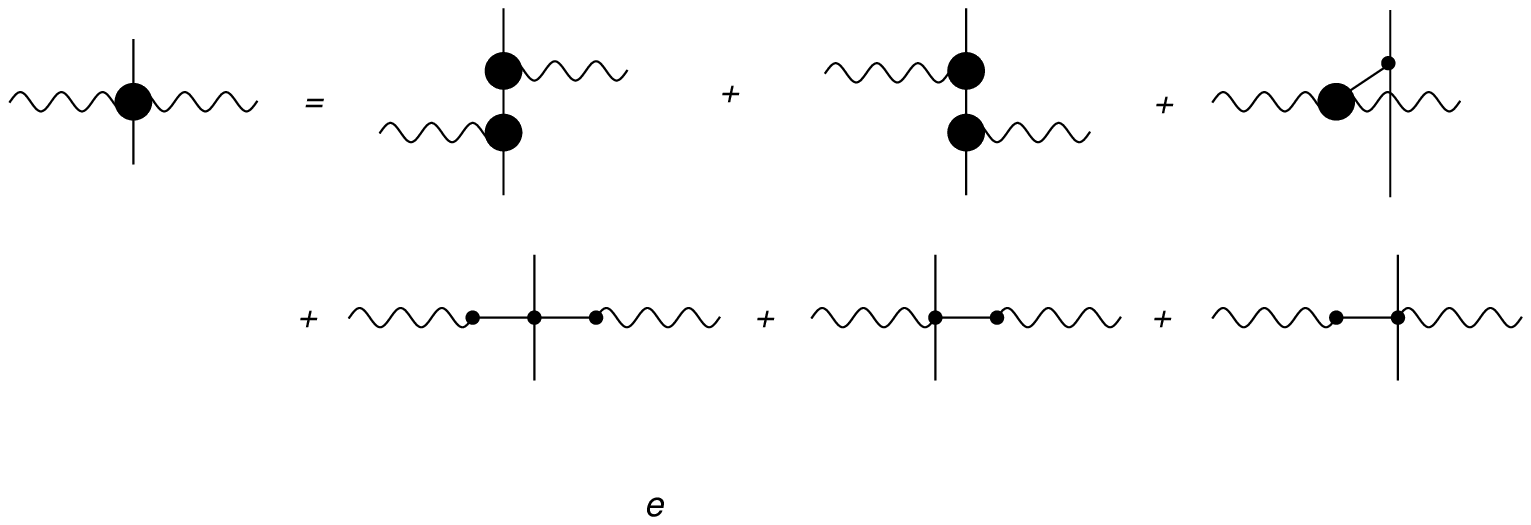,width=12cm,height=4cm}\\
Fig.3: Building blocks derived from the effective action. The straight lines denote
elementary gluons, the wavy lines correspond to reggeized gluons. Small dots stand for 
elementary vertices contained in the effective action, 
big blobs for effective vertices.       
\end{center} 
\begin{center}
\epsfig{file=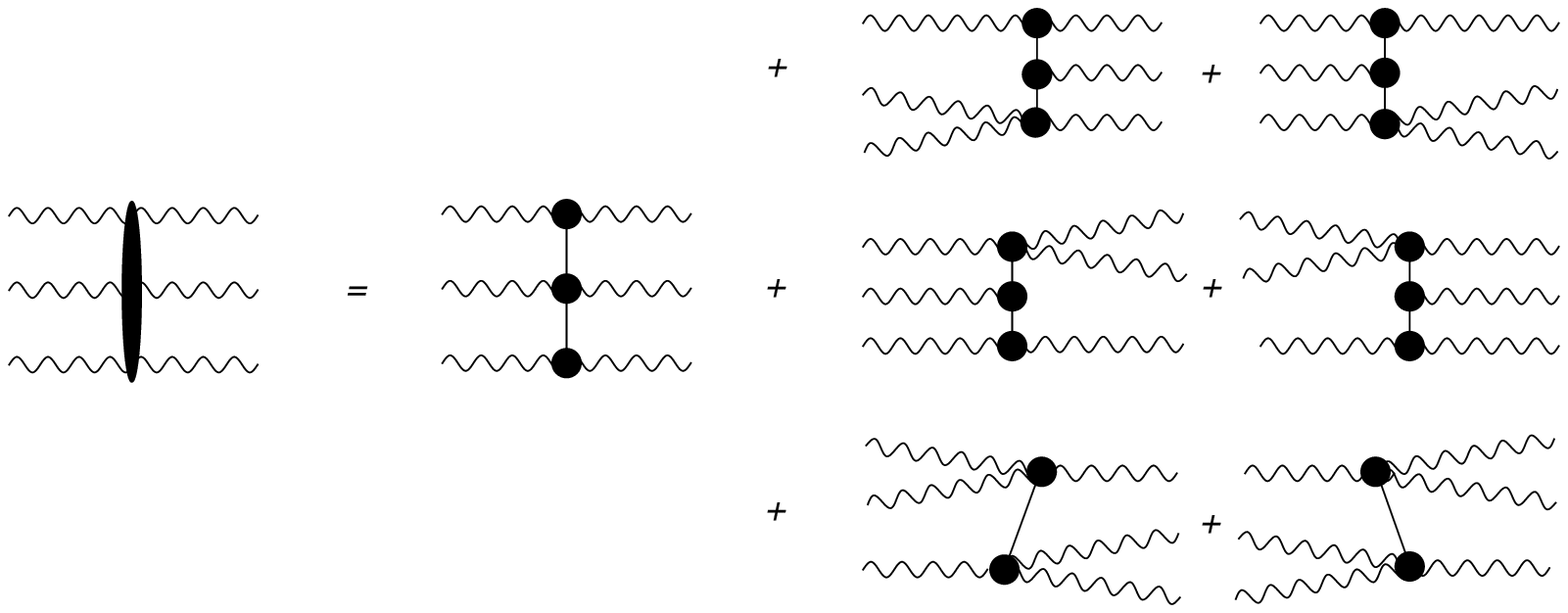,width=12cm,height=6cm}\vspace{1cm}\\
Fig.4: The full $3 \to 3$ transition kernel.\vspace{1cm}   
\end{center} 

\subsection{The longitudinal integrations}   
Our goal is the derivation of a closed expression for the $3 \to 3$ transition. We begin with the longitudinal integrations. Postponing the integrations over the $q^+$ components, we
first consider, at fixed values of $q_1^+$ and $q_3^+$, the integrations over the longitudinal components ${q'_1}^-$ and ${q'_3}^-$.  Ignoring, for the moment, the question
whether these integrations are convergent, one might be tempted to do the integrals by  
picking up poles of the $s$-channel gluon propagators. To be definite, let us consider 
the first diagram on the rhs of Fig.4, and for the RPPR vertex in the center insert 
the first term on the rhs of Fig.3e.
A quick look at the three $s$-channel propagators then shows that, for equal signs of  $q_1^+$ and $q_3^+$, the poles of the ${q'_3}^-$ integration lie on opposite sides of the integration contour, and we may pick the pole of the $k_3$ propagator  (for the 
${q'_1}^-$ integration we pick the pole of the $k_1$ propagator). In contrast, 
for opposite signs of $q_1^+$ and $q_3^+$, the two poles in ${q'_3}^-$
are on the same side of the integration contour, and the integration gives zero. 

In following we will prove that this simple picture, in fact, is correct.  
In a careful study of the longitudinal integrations we will show that only the propagator poles of the upper and lower vertical $s$-channel gluons with momenta $k_1$ and $k_3$ contribute, i.e. these gluons are on-shell. 
Since, on the rhs in Fig.4, only the first diagram has poles in both $k_1$ and $k_3$: 
we need to compute only this diagram. It consists of the product of the effective RPR production vertices above and below with the RPPR vertex in the center. In addition 
we will show that, in order to obtain a nonzero result, the $q^+$ variables have to be both positive or both negative. In deriving these results, our main problem turns out to be the convergence of the longitudinal integrations.

It is natural to start from the expressions derived from the effective action.  However,
in order to obtain the desired analytic expression, it is more convenient to 
switch to a slightly different representation. Indeed,  when addressing the integrations over the longitudinal components ${q'_1}^-$ and ${q'_3}^-$ we immediately face the
problem that, for the sum of the graphs shown in Fig.4, the integrals are divergent in the 
ultraviolet region. When taking the sum over all permutations we expect these divergences 
to cancel. However, in practice  the computation of the sum over all permutations is rather complicated. We will circumvent this problem by making use 
of the Ward identities derived in \cite{BLV}.

In \cite{BLV} it has been shown that amplitudes with reggeized gluons satisfy Ward identities.
These identities can be used to replace the unphysical polarization vectors $n^+$ and $n^-$ by
transverse momentum vectors: for each reggeized gluon with momentum $q_i$, $q'_i$ we substitute 
\bea
n^- \to -\frac{2}{q_i^+} q_{i\perp},\,\, n^+ \to - \frac{2}{{q'_i}^-}  q'_{i\perp}.
\label{Wardsubst}
\eea
This leads to the appearance of additional factors $1/q^+$ or $1/q^-$, improving the 
convergence of the integrals over longitudinal momenta. Special care has to be taken of the induced terms; details have been given in 
\cite{BLV}. We apply this procedure to the sum of all diagrams of Fig.4. 
In addition, we include the permutations over the reggeized gluons. As a result, we strip off the contributions 
containing induced vertices and multiply, 
in  the remaining QCD couplings, all horizontal lines with the corresponding transverse momenta (\ref{Wardsubst}). For the lines with momenta $q_2$ and $q'_2$ we add the induced vertices. The result is illustrated in Fig.5 (we suppress the 
permutations; we also do not show the set of diagrams obtained by left-right reflection). 
In the quartic couplings on the rhs of Fig.5, we include only those color 
structures which match the color structures of the first two terms above and below.
 \begin{center}
\epsfig{file=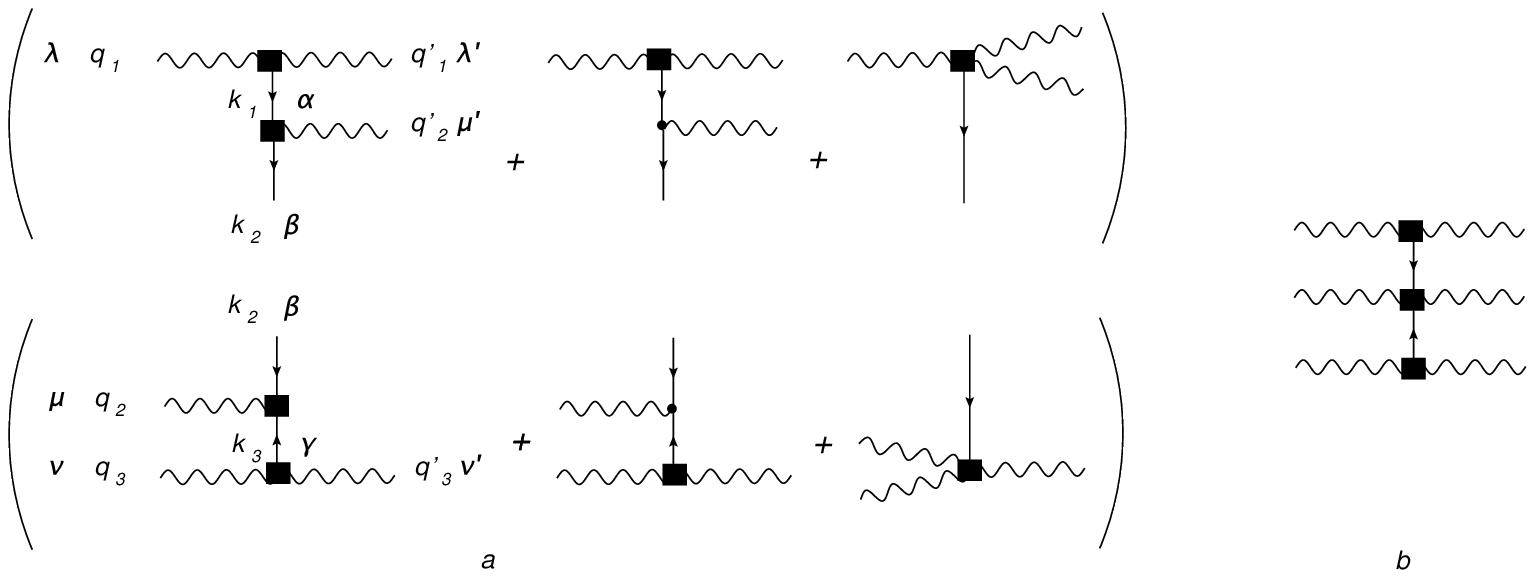,width=16cm,height=7cm}\\
Fig.5: The $3 \to 3$ kernel. Boxes denote QCD triple (quartic) gluon vertices with transverse momentum contractions on the horizontal reggeon lines,  small dots stand for induced vertices. 
\end{center}
We stress that the Ward identities are valid only for the sum of all diagrams. Compared to the original expressions - the sum of the diagrams in Fig.4 (including the permutations of the reggeized gluons and the left-right reflected counterparts) 
this use of the Ward identities is equivalent to a re-grouping of contributions. After this we again consider individual  sequences of reggeons, beginning with the planar structure $(123)$.   

In the first step we show that the integrations over ${q'_1}^-$ and ${q'_3}^-$ are ultraviolet convergent 
(at fixed transverse momenta). To this end we investigate the numerator of the sum of contributions illustrated in Fig.5. Its derivation and the final expressions are presented in the appendix.  
Here we only summarize those features which are essential for our discussion.

The sum of the contributions shown  in Fig.5. is of the form:  
\be
\label{prefactor1}
 \frac{1}{q_1^+ q_2^+ q_3^+ {q'_1}^- {q'_2}^- {q'_3}^-}\frac{N}{D_1 D_2 D_3}
\ee
with the denominators of the three $s$-channel propagators 
\bea
D_1 = k_1^2 + i \epsilon = &{q_1}^+ (-q'_1)^- + k_{1\perp}^2 + i \epsilon \\
D_2=k_2^2 +i \epsilon= &{q_1}^+ {q'_3}^- + k_{2\perp}^2 + i \epsilon \\
D_3=k_3^2+i \epsilon= &{q_3}^+ (-q'_3)^- + k_{3\perp}^2 + i \epsilon .
\eea
The numerator $N$ is a function of all longitudinal components $q_i^+$,  ${q'_i}^-$,
and the transverse momenta $q_{i\perp}$, $q'_{i\perp}$ ($i=1,2,3$). Momentum conservation implies that $q_1^+ + q_2^+ +q_3^+=0$  and ${q'_1}^- + {q'_2}^- +{q'_3}^-=0$. In (\ref{prefactor1}), the product of longitudinal components in the denominator 
results from the use of the Ward identities. 

From the results presented in the appendix we conclude that the numerator $N$
can be written as the sum of two groups of terms:
\be
N= \sum_{(ij)(kl)} q_i^+q_j^+{q'_k}^-{q'_l}^-  N^{(1)}_{(ij)(kl)}+ \sum_{(ij)} q_i^+ {q'_j}^- N^{(2)}_{(ik)},
\label{numstructure}
\ee
where the functions $N^{(1)}_{(ij)(kl)}$ and $N^{(2)}_{(ik)}$ depend upon the transverse momenta only. 
They have the important property that they vanish as any of the six transverse 
momenta $q_{i\perp}$, $q'_{i\perp}$ goes to zero.  
To investigate the ultraviolet convergence of the integrations 
in ${q'_1}^-$ and 
${q'_3}^-$, we have to combine the numerator (\ref{numstructure}) 
with the denominator and count the net powers of  ${q'_1}^-$ and  ${q'_3}^-$. 
Beginning in (\ref{numstructure}) with the first group 
(with two powers of the ${q'_i}^-$ in the numerator), 
we note that all but three pairings have (at least) two powers in  ${q'_1}^-$ 
and ${q'_3}^-$ in the denominator, either from the prefactor or from the 
propagators $D_i$. It is important to note that  all 'dangerous' terms in (\ref{numstructure}), e.g. terms proportional to ${q'_1}^-{q'_2}^-$, are found to be absent. The second group in (\ref{numstructure}) with only one power of  ${q'_i}^-$ is safe: there are always enough powers in the denominator to 
provide the convergence in the ultraviolet region. 

In the next step we investigate the infrared region of the integrations over  ${q'_1}^-$ and ${q'_3}^-$. At first sight one might think that there are poles at ${q'_i}^-=0$ which
need to be regularized. However, these poles are spurious and cancel, once we 
take the sum of the six permutations over the reggeized gluons on the rhs (for any fixed sequence of gluons on the lhs) and perform the integration over the $q^+$ components. To see this in detail, we make use of the fact that the impact factors in Fig.2 
are symmetric under the exchange of the momenta of the reggeized gluons. 
This implies that in (\ref{prefactor1}) the transverse momenta inside 
the numerator $N$ and the denominators $D_i$ can be written in the   
same form for all permutations, and the permutations only differ in their 
dependence upon the longitudinal components. It is then straightforward to show
that the poles at  ${q'_i}^-=0$ have zero residues. As an example, 
in the first group of (\ref{numstructure}) consider the term
\be
\int \frac {d{q'_1}^- d{q'_3}^-}{{q'_1}^-}\int \frac{dq_1^+ dq_3^+}{q_1^+} \frac{1}{D_1D_2D_3}.
\ee
We add the permutation with gluon "2"   and gluon "3"  being interchanged and 
use the explicit form of the propagators:  
\be
\int \frac {d{q'_1}^- d{q'_3}^-}{{q'_1}^-}\cdot \\
\ee
\bea
\cdot \int \frac{dq_1^+ dq_3^+}{q_1^+} 
\frac{1}{{q_1}^+ (-q'_1)^- + k_{1\perp}^2)} 
\left(   \frac{1}{   ({q_1}^+ {q'_3}^- + k_{2\perp}^2)
                        ({q_3}^+ (-q'_3)^- + k_{3\perp}^2) } 
                        \right. \nonumber\\ \left. 
+ \frac{1}{   ({q_1}^+ {q'_2}^- + k_{2\perp}^2)
                        ({q_3}^+ (-q'_2)^- + k_{3\perp}^2)  }  \right). \nonumber
\eea
In the limit ${q'_1}^- \to 0$ the sum of the second and third lines vanishes: for ${q'_1}^- \to 0$ we have ${q'_2}^- \to 
- {q'_3}^-$, and by the change $q_1^+ \to - q_1^+$, $q_3^+ \to - q_3^+$, the second 
fraction equals the first one, up to a minus sign.  A similar argument holds for all the other 
terms of the first group in (\ref{numstructure}): one always finds  
pairs of two permutations in which the pole at ${q'_i}^-=0$ cancels.  As to the second group, one first shows, 
in exactly the same way, that there are no single poles in any ${q'_i}^-$. Finally, by adding all six permutations 
one easily sees that also the regions ${q'_i}^-=0$, ${q'_j}^-=0$ do not lead to any singularity, i.e. we have no singularities 
from the denominators $1/{q'_i}^- {q'_j}^-$ ($i \neq j$).  Having shown that in the sum over all six permutations there 
are no infrared singularities, we are allowed to add, in all denominators  of $1/{q'_i}^-$ 
small imaginary parts as intermediate regulators: $1/({q'_i}^- \pm i \epsilon)$. 
The prescription has to be the same for all permutations, and the final answer, of course,  will not depend upon which prescription we have chosen. We find it convenient to 
use the principal value prescription, i.e. in (\ref{prefactor1}) we substitute
\be
\int \frac{d{q'_i}^-}{{q'_i}^-} \to {\cal P}\int \frac{d{q'_i}^-}{{q'_i}^-} = 
\int \frac{ d{q'_i}^-}{2}\left(  \frac{1}{{q'_i}^- +i \epsilon}+ \frac{1}{{q'_i}^- - i \epsilon}
\right) 
\ee    
for each denominator.
 
In the third part of our argument we perform, for each permutation separately,  the integrations over  ${q'_1}^-$ and ${q'_3}^-$ (at fixed values of $q_1^+$ and $q_3^+$). 
Detailed expressions are presented in the appendix, and we only quote the most important 
results: denoting the four regions ($q_1^+>0,\, q_3^+>0$), ($q_1^+<0,\, q_3^+<0$), ($q_1^+>0,\, q_3^+<0$), and ($q_1^+<0,\, q_3^+>0$) by $>>$, $<<$, $><$, $<>$, reps., 
we find: \\
(i) only the equal sign regions, $<<$ and $>>$, contribute\\
(ii) apart from the overall factor $1/2$, the results are the same as obtained from taking residues of the 
poles coming from $D_1$ and $D_3$, i.e. by putting  the lines $k_1$ and $k_3$ on-shell. The factor $1/2$ is removed by 
summing over the two regions $>>$ and $<<$: in the following
 we restrict the longitudinal variables $q_1^+$ and $q_3^+$ to be positive.   

This completes the derivation of the result stated at the beginning of this subsection:
in Fig.1  the upper and lower vertical $s$-channel gluons with momenta $k_1$ and $k_3$ are on-shell, and the longitudinal $'+'$ components of $k_1$ and $k_3$ must have equal
sign. 
We could therefore use our expressions of the RPR and RPPR vertices, given in the appendix, with on-shell s-channel gluons, and combine them with the results of 
the integrations over the ${q'_i}^-$ components. 

Nevertheless, we find it convenient to return to Fig.4, i.e. to the expressions for the kernel before our use of the Ward identities.  The reason is that, in the integrals $I_2$, $I_3$, and $I_4$ and $I_6$ one finds the 
denominator $D_{13}$ which resulted from inserting the pole values for 
${q'_1}^-$ and ${q'_3}^-$ from $D_1$ and $D_3$ into the denominator 
$-1/{q'_2}^-= 1/({q'_1}^- +{q'_3}^-)$. In the subsequent integration over the longitudinal 
$q_i^+$ variables the presence of this denominator will lead to unwanted poles, e.g. of the form $1/(k_{1\perp}^2 - k_{3\perp}^2)$,  
which in the final result will cancel. In order to avoid this unnecessary 
complication it is more convenient to return to Fig.4. On the right hand side of Fig.4, only the first diagram
has the two $s$-channel propagator poles in $D_1$ and $D_3$, and all the other 
diagrams can be disregarded.

\subsection{Final expression for the $3 \to 3$ vertex}
After this rather long 'detour' we address the final step and compute the first diagram on the right hand side  of Fig.4, with 
on-shell $s$-channel gluons with momenta $k_1$ and $k_3$. Beginning with the RPPR vertex in the center (Fig.3e), it is convenient 
to make use of the existing two gluon production vertex in quasi-multi-Regge  
kinematics. This vertex has been calculated before, 
in the context of the NLO calculation of the BFKL kernel \cite{2gluon}. 
A convenient derivation starting from the effective action has been described in 
\cite{Antonov:2004hh}. For our 
discussion we start from the expression given in \cite{Fadin:2003xs}. Denoting the 
momenta of the two produced $s$-channel gluons by $k_1$ and $k_3$, and  
turning them into incoming momenta, we arrive at the tensor:
\bea   
b^{\alpha \beta}(x_1) = - {\q_2}^{\,2} \frac{k_1^{\alpha} k_3^{\beta}}{\vk_1^2 \vk_3^2}-
\hspace{10cm}\nonumber\\
-\frac{g^{\alpha \beta}}{2} \frac{x_1x_3 (\q_2^2 +2\q_2\vk_1)}{\sigma_1}  
-\frac{x_3k_1^{\alpha}q_2^{\beta} + x_1 q_2^{\alpha} (q_2+k_1)^{\beta}}{\sigma_1}
-\frac{x_1 \q_2^2 k_1^{\alpha} (q_2+k_1)^{\beta}}{\vk_1^2 \sigma_1}
\nonumber \\
-\frac{g^{\alpha \beta}}{2} \frac{x_1x_3 (\q_2^2 +2\q_2\vk_3)}{\sigma_3}  
-\frac{x_1q_2^{\alpha}k_3^{\beta} + x_3 (q_2+k_3)^{\alpha}q_2^{\beta}}{\sigma_3}
-\frac{x_3 \q_2^2 k_3^{\beta}(q_2+k_3)^{\alpha}}{\vk_3^2 \sigma_3},
\label{3to3tensor}
\eea                
where the second line belongs to the first diagram in Fig.3e, the third line to 
the second diagram, and we have introduced $k_{i\perp}^2 = - \vk_i^2$.
The denominators in (\ref{3to3tensor}) are defined as
\bea 
\sigma_1=x_1 (\q_2+\vk_1)^2+x_3 \vk_1^2 \nonumber\\
\sigma_3=x_3 (\q_2+\vk_3)^2+x_1 \vk_3^2.
\label{denominators}
\eea        
We have rescaled our $q_i^+$ variables (both $q_1^+$ and $q_2^+$ are positive): 
\be
q_i^+ = x_i\, q^+,\,\,i=1,3;\,\,q^+=q_1^+ + q_3^+.
\label{scaling}
\ee
The integral over $q^+$ yields the logarithm of the energy corresponding to the BFKL evolution, whereas the integration over $x_1$ is a part of the RPPR vertex. 

This expression has to be integrated over $x_1$ with the integration measure
$1/\left(x_1(1-x_1)\right)$. The integral diverges logarithmically at $x_1=1$ 
and at $x_1=0$. To be specific, let us consider Fig.5 (which corresponds to the first diagram in Fig.3e):   
at $x_1=0$ we have $q_3^+ \gg q_1^+$, i.e. we are in the 
kinematical region in which the gluons with momenta $k_1$ and $k_3$ are separated by a large rapidity 
interval, and the gluon with momentum $k_2=k_1-q'_2$  becomes a $t$-channel reggeized gluon having negligible 
longitudinal momenta. This 
configuration belongs to the leading order BKP evolution and, in order to find the 
next-to-leading order result, we have to subtract this contribution. A similar procedure holds for 
the limit $x_1 \to 0$. For the NLO result we impose the intermediate rapidity cut-off:
\be
|y_1 - y_3| < \Delta,
\label{rapcut}
\ee  
where $y_i= \ln (q_i^+/|\vk_i|)$ ($i=1,3$) denotes the 
rapidity of gluon $i$. Including the subtraction of the LO contributions 
proportional to $b^{\alpha \beta}(0) + 
b^{\alpha \beta}(1)$ we find the following regularization prescription:
\bea
\int_0^1 \frac{dx_1}{x_1 (1-x_1)} b^{\alpha \beta}(x_1)| _+ = 
\int_{e^{-\Delta} \frac{|\vk_1|}{|\vk_3|}}^{1- e^{-\Delta} \frac{|\vk_3|}{|\vk_1|}}
\frac{dx_1}{x_1(1-x_1)} b^{\alpha \beta}(x_1) - 
\Delta \left( b^{\alpha \beta}(0) + b^{\alpha \beta}(1) \right)
\nonumber\\
= \int_0^1 \frac{dx_1}{x_1} \left (b^{\alpha \beta}(x_1)-b^{\alpha \beta}(0) \right)
+ \int_0^1 \frac{dx_1}{1-x_1} \left(b^{\alpha \beta}(x_1)-b^{\alpha \beta}(1)\right)
\nonumber\\
+ b^{\alpha \beta}(0) \ln \frac{|\vk_3|}{|\vk_1|}  + b^{\alpha \beta}(1) \ln \frac {|\vk_1|}{|\vk_3|}, 
\label{regularization}
\eea
where, in the second line, we have taken the limit $\Delta \to \infty$. 
The integration over $x_1$ leads to the following expression for the RRPP vertex:  .
\bea
V^{\alpha \beta}=
\ln \frac{(\q_2+\vk_1)^2}{\sqrt{\vk_1^2 \vk_3^2}}
\left(-\frac{g^{\alpha \beta}}{2} +
\frac{k_1^{\alpha} q_2^{\beta}}{\vk_1^2} - \frac{\q_2^2}{(\q_2+\vk_1)^2}
\frac{k_1^{\alpha} (q_2+k_1)^{\beta}}{\vk_1^2}- 
\frac{q_2^{\alpha} (q_2+k_1)^{\beta}}{(\q_2+\vk_1)^2} \right) +\nonumber \\
 + \ln \frac{(\q_2+\vk_3)^2}{\sqrt{\vk_1^2 \vk_3^2}}
\left(-\frac{g^{\alpha \beta}}{2} +
\frac{q_2^{\alpha} k_3^{\beta}}{\vk_3^2} - \frac{\q_2^2}{(\q_2+\vk_3)^2}
\frac{(q_2+k_3)^{\alpha} k_3^{\beta}}{\vk_3^2}- 
\frac{(q_2+k_3)^{\alpha} q_2^{\beta}}{(\q_2+\vk_3)^2} \right).    
\label{RRPPtensor}
\eea
Including the contraction with the effective vertices from above and below 
we can write the result in the simplified form:
\bea
\label{K123}
K_{123}&=& \left(\frac{q_1^{\alpha}}{\q_1^2} - \frac{k_1^{\alpha}}{\vk_1^2}\right) 
V_{\alpha \beta}
\left(\frac{q_3^{\beta}}{\q_3^2} - \frac{k_3^{\beta}}{\vk_3^2}\right)\nonumber\\
&=& -  
\left(\frac{q_1^{\alpha}}{\q_1^2} - \frac{k_1^{\alpha}}{\vk_1^2}\right)
\Big[
\ln \frac{(\q_2+\vk_1)^2}{\sqrt{\vk_1^2 \vk_3^2}} 
T_{\alpha \beta}(k_1)T_{\beta \gamma}(q_2+k_1) + \\
&&\hspace{3cm} + \ln \frac{(\q_2+ \vk_3)^2}{\sqrt{\vk_1^2 \vk_3^2}} 
T_{\alpha \beta}(q_2+k_3)T_{\beta \gamma}(k_3) \Big] 
\left(\frac{q_3^{\gamma}}{\q_3^2} - \frac{k_3^{\gamma}}{\vk_3^2}\right),\nonumber
\eea
where 
\be
T_{\alpha \beta}(k) = g_{\alpha \beta} - 2 \frac{k_{\alpha} k_{\beta}}{\vk^2}.
\label{gaugechange}
\ee
In (\ref{K123}) we have included the propagators for the two gluons with momenta 
$q_1$ and $q_3$. 
For later convenience it is useful to present the vertex also in complex
notation. Introducing $q=q_x+iq_y$, $q^*=q_x -iq_y$ etc. we find:
\bea
\tilde{K}_{123}=-\frac{1}{4} \frac{1}{q_2 q_2^*} \Biggl\{ 
\log\frac{|q_2+k_1|^2}{|k_1||k_3|}
\left[ \frac{q'^{*}_1 q'_3}{q_1^* q_3}\frac{1}{k_1 k_3}
\frac{(q_2+k_1)}{(q_2+k_1)^*}+ {\rm c.c.} \right]
+ \nonumber\\
\log\frac{|q_2+k_3|^2}{|k_1| |k_3|}
\left[ \frac{q'_1 q'^{*}_3}{q_1 q_3^*}\frac{1}{k_1 k_3}
\frac{(q_2+k_3)}{(q_2+k_3)^*}+ {\rm c.c.} \right]
\Biggr\}.
\label{tK123}
\eea
Here we have included also the propagator of the  gluon with momentum $q_2$. 

Let us mention a few important properties of this $3\to3$ vertex.
First we note that the poles from the propagators of the $s$-channel gluon with momentum 
$k_2$ in (\ref{RRPPtensor}) cancel: this 
is seen most directly in the complex representation (\ref{tK123}).
Next we note that the vertex, after removing the propagators on the left side of the $3\to3$ vertex, 
vanishes as any of the external momenta $\q_i$, $\q'_i$ goes to zero, in agreement with 
the arguments of section 3.1 based on the Ward identities.  

This completes our derivation of the $3 \to3$ vertex. Together with the NLO 
$2\to2$ kernel presented in section 2, it constitutes the NLO kernel of the BKP evolution equation.
 
\section{Conclusions}

In this paper we have presented analytic expressions of the NLO kernel of the 
BKP evolution equations in the Odderon channel, consisting of the NLO 
corrections of the $2\to2$ BFKL kernel in the symmetric color octet channel and of the Born approximation of the connected $3\to3$ transition kernel. In deriving the final form of the 
$3 \to 3$ (eqs.(\ref{K123}) and (\ref{tK123})) we found it helpful to make use 
of Ward identities of amplitudes of reggeized gluons which have been derived recently.
In \cite{BLV} these Ward identities had been used for obtaining a new derivation   
of the BFKL kernel; in the present paper we have derived, by the same methods,  
compact expressions for the RPRR, RRPR, and RPPR vertices.  
Starting from the results of  the $2\to2$ and 
$3\to3$ transition kernels it will be possible to calculate the NLO intercept of the Odderon solution. Work along these lines is in progress.     

Our NLO calculation of the $2\to2$ kernel also applies to the large-$N_c$ 
limit of the C-even and odd BKP equations of $n\ge3$ reggeized gluons in the color singlet state. 
In the (planar) large-$N_c$ limit any pair of neighboring reggeized gluons is in an 
octet state, and the gluon trajectories of the $n$ gluon lines can be 
distributed among the $2 \to2 $ kernels in such way that the kernels are infrared finite. Our results can also be used for the large $N_c$ limit of the BKP-equations for n reggeized gluons in the adjoint representation. This case has attracted interest in the context of studies of the  
AdS/CFT duality, in particular for the computation of the remainder function.   
Compared to the  color singlet case, the infrared finiteness of the 
$2 \to2$ kernels  now works in a slightly different way. Labeling the reggeized gluons of the n-gluon state by $1,...,n$ with momenta 
$q_1$, ... $q_n$, we distribute the gluon trajectories in the same way as in the singlet case. But since in the planar approximation of the adjoint representation, when compared to the color singlet configuration which lives on the surface of a cylinder, one $2 \to 2$ kernel is missing  (namely the interaction between reggeons $n$ and $1$), the compensation of infrared singularities is incomplete. In order to achieve infrared finiteness, we subtract the gluon trajectory as a function the sum of all transverse 
momenta, $q=q_1+...+q_n$. This subtraction is in agreement with the fact that, in the 
BDS formula, this trajectory is contained already in the BDS part and thus should not be 
part of the remainder function. In this way, also the color octet case is infrared finite, 
and should have the dual conformal invariance in momentum space, observed 
in \cite{Bartels:2008sc} and \cite{Lipatov:2009nt}.

Our derivation of the $3\to3$ transition kernel, so far, is restricted to 
the system of three reggeized gluons in the Odderon channel. Its use for 
the C-even odd-signature three gluon contribution to the spin structure 
function $g_2$ \cite{Lipatov:1999pv}, as well as for the  large-$N_c$ limit of the C-even and odd BKP equations 
of $n$ reggeized gluons in the color singlet state requires 
a few additional steps which will be presented in a future paper;
the same applies also to its use for the remainder function of the $2 \to n$ amplitude 
within the AdS/CFT duality.  In a future paper we hope to come back to these generalizations.\\[1cm] 
{\bf Acknowledements}:
This work has been supported by the SFB 676 'Particles, Strings and the 
Early Universe: the structure of Matter and Space-Time'.
Three of us (V.S.Fadin, L..Lipatov, and G.P.Vacca) express their gratitude 
for the hospitality of DESY and of the II.Institut of Theoretical Physics,
Hamburg University. One of us (J.Bartels)
thanks the INFN and the University of Bologna for their hospitality.      

\appendix

\setcounter{equation}{0}

\renewcommand{\theequation}{A.\arabic{equation}}
\section{Appendix}

In this appendix we derive the numerator of the sum of diagrams shown in Fig.5.
Let us first describe the building blocks. Beginning with the left hand part of Fig.5a, the production vertices at the upper and  at the lower end 
(black boxes) are obtained from the effective production vertices illustrated in the first line of Fig.3 by removing the induced contributions. What is left are the QCD triple gluon vertices, contracted with transverse momenta from the left and from the right hand side
(see eq.(\ref{Wardsubst})).   
Below the upper production vertex we have the left central vertex and the right central vertex: they are obtained from the effective vertices in the second line of Fig.3 by disregarding the induced vertices and by multiplying with the corresponding transverse 
momenta: this leads to the QCD triple gluon vertices with a transverse momentum contraction for the reggeized gluon. At the end, we have to re-add the induced terms for the gluons with momenta $q_2$ and $q'_2$.

 \subsection{Building blocks}
We collect the building blocks. The overall multiplicative factor is  
\be
\label{prefactor}
\frac{1}{D_1 D_2 D_3} \frac{2^6}{q_1^+ q_2^+ q_3^+ {q'_1}^- {q'_2}^- {q'_3}^-}
\ee
with
\bea
D_1 = k_1^2 + i \epsilon = &{q_1}^+ (-q'_1)^- + k_{1\perp}^2 + i \epsilon \\
D_2=k_2^2 +i \epsilon= &{q_1}^+ {q'_3}^- + k_{2\perp}^2 + i \epsilon \\
D_3=k_3^2+i \epsilon= &{q_3}^+ (-q'_3)^- + k_{3\perp}^2 + i \epsilon. 
\eea

Next we list the four vertices (black boxes) in the left column of Fig.5a. For the QCD triple gluon vertex 
\begin{center}
\epsfig{file=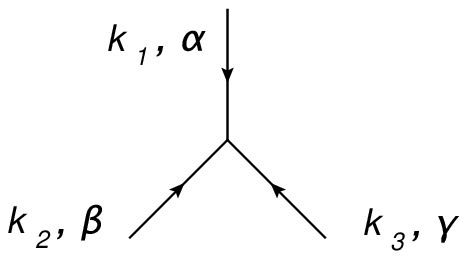,width=3cm,height=2cm}\\
\end{center}
we use the notation
\be
\gamma^{\alpha \beta \gamma}(k_1,-k_3) = g^{\alpha \beta} (k_1-k_2)^{\gamma}
+g^{\beta \gamma} (k_2-k_3)^{\alpha} +g^{\gamma \alpha} (k_3-k_1)^{\beta}. 
\ee 
For the first term, the upper vertex, we have from \cite{BLV}:
\bea
 \left(   (q_1)_{\perp \lambda} \gamma^{\lambda \alpha \lambda'}(q_1,q'_1) (q'_1)_{\perp \lambda'} \right) =
 \nonumber\\  (q_{1\perp})^{\alpha} (q_1-q'_1)_{\perp} q'_{1\perp} - (q'_{1\perp})^{\alpha} (q_1-q'_1)_{\perp} q_{1\perp} - \left( \frac{(n^+)^{\alpha}}{2} (q'_1)^- + \frac{(n^-)^{\alpha}}{2} (q_1)^+\right)  q_{1\perp}q'_{1\perp}.
 \label{uppertgt}
\eea
For the left central vertex for the gluon with momentum $q'_2$ we find:
\bea
 \gamma^{\beta \alpha  \mu'}(-k_2, q'_2) (q'_{2\perp}) _{\mu'}
 = - g^{\alpha \beta} \,\,q'_{2\perp}(k_2+k_1)_{\perp} \hspace{2cm} \nonumber\\
  + \frac{(n^+)^{\beta}}{2} (q'_{2\perp})^{\alpha}    (-q'_1+q'_2)^-
      + \frac{(n^-)^{\beta}}{2}(q'_{2\perp})^{\alpha} q_1^+
      + \frac{(n^+)^{\alpha}}{2} (q'_{2\perp})^{\beta} (-q'_2+q'_3)^-
      + \frac{(n^-)^{\alpha}}{2} (q'_{2\perp})^{\beta} q_1^+ \nonumber\\
     + (q'_{2\perp})^{\alpha}(k_1+q'_2)_{\perp}^{\beta} + (q'_{2\perp})^{\beta}(-q'_2+k_2)_{\perp}^{\alpha} ,\hspace{2cm}
\eea
which we can also write in the form:
\bea
 \gamma^{\beta \alpha \mu'}(-k_2, q'_2) (q'_{2\perp}) _{\mu'}
 =\nonumber \\
  - g^{\alpha \beta} \,\,q'_{2\perp}(k_2+k_1)_{\perp} + (q'_{2\perp})^{\beta} k_1^{\alpha} 
 +k_2^{\beta} (q'_{2\perp})^{\alpha} -{q'_2}^- \left( (q'_{2\perp})^{\beta} (n^+)^{\alpha}
 - (n^+)^{\beta} (q'_{2\perp})^{\alpha} \right).
 \label{rightgt}
\eea
For the induced term we have:
\bea
- \frac{{q'_2}^-}{2} \frac{{q'}_{2\perp}^2}{k_1^+}    (n^+)^{\alpha} (n^+)^{\beta} =- \frac{{q'}_{2\perp}^2}{2}     \frac{{q'_2}^-}{{q_1}^+}    (n^+)^{\alpha} (n^+)^{\beta}.
\label{inducedq_2prime}
\eea
Similarly, the right central vertex for the gluon with momentum $q_2$ reads:
 \bea
(q_{2\perp}) _{\mu}  \gamma^{\mu \beta  \gamma}(q_2,-k_3)
 = g^{\beta\gamma} \,\,q_{2\perp}(k_2-k_3)_{\perp} \hspace{2cm} \nonumber\\
  + \frac{(n^+)^{\gamma}}{2} (q_{2\perp})^{\beta}    (-q'_3)^-
      + \frac{(n^-)^{\gamma}}{2}(q_{2\perp})^{\beta} (q_2-q_1)^+
      + \frac{(n^+)^{\beta}}{2} (q_{2\perp})^{\gamma} (-q'_3)^-
      + \frac{(n^-)^{\beta}}{2} (q_{2\perp})^{\gamma} (q_3-q_2)^+ \nonumber\\
     + (q_{2\perp})^{\beta}(q_2-k_2)_{\perp}^{\gamma} + (q_{2\perp})^{\gamma}(k_3-q_2)_{\perp}^{\beta}, \hspace{2cm}
\eea
for  which we also have:
\bea
(q_{2\perp}) _{\mu}  \gamma^{\mu \beta \gamma}(q_2,-k_3)\nonumber\\
 = g^{\beta\gamma} \,\,q_{2\perp}(k_2-k_3)_{\perp} + (q_{2\perp})^{\beta} k_3^{\gamma} 
 -k_2^{\beta} (q_{2\perp})^{\gamma} +q_2^+ \left( (q_{2\perp})^{\beta} (n^-)^{\gamma}
 - (n^-)^{\beta} (q_{2\perp})^{\gamma} \right) .
 \label{lefttg}
\eea
The induced term for this  reggeized gluon is:
\bea
- \frac{(q_2)^+}{2} \frac{q_{2\perp}^2}{k_3^-}    (n^-)^{\beta} (n^-)^{\gamma} = 
\frac{q_{2\perp}^2}{2}  \frac{(q_2)^+}{(q'_3)^-}  (n^-)^{\beta} (n^-)^{\gamma}.
\label{inducedq_2}
\eea
Finally the lower vertex:
\bea
 \left(   (q_3)_{\perp \nu} \gamma^{\nu \gamma \nu'}(q_3,-q'_3) (q'_3)_{\perp \nu'} \right) =\nonumber\\  (q_{3\perp})^{\gamma} (q_3-q'_3)_{\perp} q'_{3\perp} - (q'_{3\perp})^{\gamma} (q_3-q'_3)_{\perp} q_{3\perp} - \left( \frac{(n^+)^{\gamma}}{2} (q'_3)^- + \frac{(n^-)^{\gamma}}{2} (q_3)^+\right)  q_{3\perp}q'_{3\perp}.
\label{lowertgt}
\eea
For the upper line in Fig.5a, the contribution of the quartic coupling is: 
\be
(q_{1\perp})^{\beta} q'_{2\perp}q'_{1\perp} - (q'_{1\perp})^{\beta} q'_{2\perp} q_{1\perp}.
\ee
Similarly for the lower line: 
\bea
(q_{3\perp})^{\beta} q_{2\perp}q'_{3\perp} - (q'_{3\perp})^{\beta} q_{2\perp} q_{3\perp}.
\label{lowerquartic}
\eea
Finally, the central part of Fig.5b is given by: 
\be
g^{\alpha\gamma} q_{2\perp}q'_{2\perp} - (q_{2\perp})^{\alpha} (q'_{2\perp})^{\gamma}.
\label{centralquartic}
\ee
\subsection{The effective RPRR and RRPR vertices}
  
Next we write down the explicit expression for the lower RRPR vertex (lower line in Fig.5a). Combining (\ref{lefttg}), (\ref{lowertgt}), (\ref{inducedq_2}), and (\ref{lowerquartic}) we obtain:
\bea
-q_2^+ {q'_3}^- q_{2\perp}^{\beta}\; q_{3\perp}q'_{3\perp}
- k_2^{\beta} (q_{2\perp}\times k_{3\perp}) (q_{3\perp}\times q'_{3\perp})+
(k_3^+ k_3^- + k_{3\perp}^2) (n_{\perp}^{\beta} \times q_{2\perp})(q_{3\perp}\times q'_{3\perp})
\nonumber\\
-\left(\frac{(n^+)^{\beta}}{2} (q'_3)^-  + \frac{(n^-)^{\beta}}{2} (q_3)^+\right)\; q_{3\perp}q'_{3\perp} \; q_{2\perp}(k_{2\perp}-k_{3\perp}) - (q_2)^+ (n^-)^{\beta} \;(q_{2\perp}\times k_{3\perp}) (q_{3\perp} \times q'_{3\perp})\nonumber \\
+ (n_{\perp}^{\beta}\times k_{3\perp})\;(q_{3\perp}\times q'_{3\perp})\; q_{2\perp}(k_{2\perp}-k_{3\perp}) - \frac{(n^-)^{\beta}}{2}\; (q_2)^+ \;q_{2\perp}^2 \;q_{3\perp}q'_{3\perp}, \hspace{1cm}
\eea
where we have introduced the vector product-like notation $(a\times b)= a_xb_y-a_yb_x$ and defined the unit vector $n_{\perp}^{\beta}$ with $n_{\perp}^{\beta} q_{\perp}=  (q_{\perp})^{\beta}$.
Alternatively we can write:  
\bea
\frac{(n^-)^{\beta}}{2}\big[ -(q_3)^+ \;q_{3\perp}q'_{3\perp}  q_{2\perp}(k_{2\perp}-k_{3\perp})+ (q_3-q_2)^+ (q_{2\perp}\times k_{3\perp})(q_{3\perp}\times q'_{3\perp})  - (q_2)^+ q_2^2\; q_{3\perp}q'_{3\perp}\big]\nonumber\\
+\frac{(n^+)^{\beta}}{2} \big[-(q'_3)^- \;q_{3\perp}q'_{3\perp} q_{2\perp}(k_{2\perp}-k_{3\perp}) - (q'_{3})^- (q_{2\perp}\times k_{3\perp})(q_{3\perp}\times q'_{3\perp})  \big]\nonumber\\
-q_2^+ {q'_3} ^-q_{2\perp}^{\beta} q_{3\perp}q'_{3\perp} +(-q_3^+ {q'_3}^- +k_{3\perp}^2)  (n_{\perp}^{\beta}\times q_{2\perp})(q_{3\perp}\times q'_{3\perp}) 
\nonumber\\
 - k_{2\perp}^{\beta} (q_{2\perp}\times k_{3\perp}) (q_{3\perp}\times q'_{3\perp}) + (n_{\perp}^{\beta}\times k_{3\perp})(q_{3\perp}\times q'_{3\perp}) \;q_{2\perp}(k_{2\perp}-k_{3\perp}).\hspace{1cm}
\label{lowerRRPR}
\eea
One verifies that  (\ref{lowerRRPR}) satisfies the Ward identity,
i.e. after contracting with $(k_2)_{\beta}$ we obtain zero. We also note that if we put
the $k_2$-line on mass shell and multiply with a physical (transverse) polarization vector,
the induced term of the reggeized gluon with momentum $q_2$ does not contribute.

An analogous expression is obtained for the upper RPRR vertex (upper line in Fig.5a):  
\bea
\frac{(n^+)^{\beta}}{2}\big[ (q'_1)^- \;q_{1\perp}q'_{1\perp}  q'_{2\perp}(k_{1\perp}+k_{2\perp})+ (q'_2-q'_1)^- (q'_{2\perp}\times k_{1\perp})(q_{1\perp}\times q'_{1\perp})  +(q'_2)^- {q'_{2\perp}}^2\; q_{1\perp}q'_{1\perp}\big]
\nonumber\\
+ \frac{(n^-)^{\beta}}{2}\big[(q_1)^+ \;q_{1\perp}q'_{1\perp} q'_{2\perp}(k_{1\perp}+ k_{2\perp}) + (q_{1})^+ (q'_{2\perp}\times k_{1\perp})(q_{1\perp}\times q'_{1\perp})  \big]    \nonumber\\
+{q'_2}^- q_1^+ (q'_{2\perp})^{\beta} q_{1\perp}q'_{1\perp} +(-q_1^+ {q'_1}^-+ k_{1\perp}^2)
(n_{\perp}^{\beta}\times q'_{2\perp}) (q_{1\perp}\times q'_{1\perp})\nonumber\\
+k_{2\perp}^{\beta}\;(q'_{2\perp} \times k_{1\perp}) (q_{1\perp}\times q'_{1\perp}) 
-(n_{\perp}^{\beta}\times k_{1\perp}) (q_{1\perp} \times q'_{1\perp})\; 
q'_{2\perp}(k_{1\perp}+k_{2\perp}).\hspace{1cm}
\label{upperRPRR}
\eea

The contribution of the quartic coupling in Fig.5b reads:
\bea
\left(q_1^+ {q'_3}^- +k_{2\perp}^2 \right) \Big[ \frac{1}{2} \left( q_3^+ {q'_1}^- + q_1^+ {q'_3}^-\right)
q_{1\perp}q'_{1\perp} \; q_{2\perp} q'_{2\perp} á,q_{3\perp} q'_{3\perp} +\nonumber \\  
k_{1\perp} q'_{2\perp}\; k_{3\perp} q_{2\perp}\; (q_{1\perp}\times q'_{1\perp})(q_{3\perp} \times q'_{3\perp}) \Big].
\eea

For later purposes we remind that the production vertex  in (\ref{uppertgt}) 
satisfies the Ward identity, i.e. it vanishes when contracted with $k_1$.
The same applies to the lower production vertex (\ref{lowertgt}) and its contraction with $k_3$, resp.. 

\subsection{The effective RPPR vertex}

Furthermore, from the building blocks listed in subsection A.1 we can form the gauge invariant RPPR vertex (Fig.6). 
To this end we combine, in Fig.6a, the left and right central vertices, (\ref{rightgt}) and (\ref{lefttg}) 
(together with their induced terms). In Fig.6b we collect the mirror reflected 
diagrams of Fig.6a. Fig.6c shows the nonplanar diagram (which in the odderon case drops out, 
once we convolute with the impact factors),
and Fig.6d shows the quartic coupling (\ref{centralquartic}).
\begin{center}
\epsfig{file=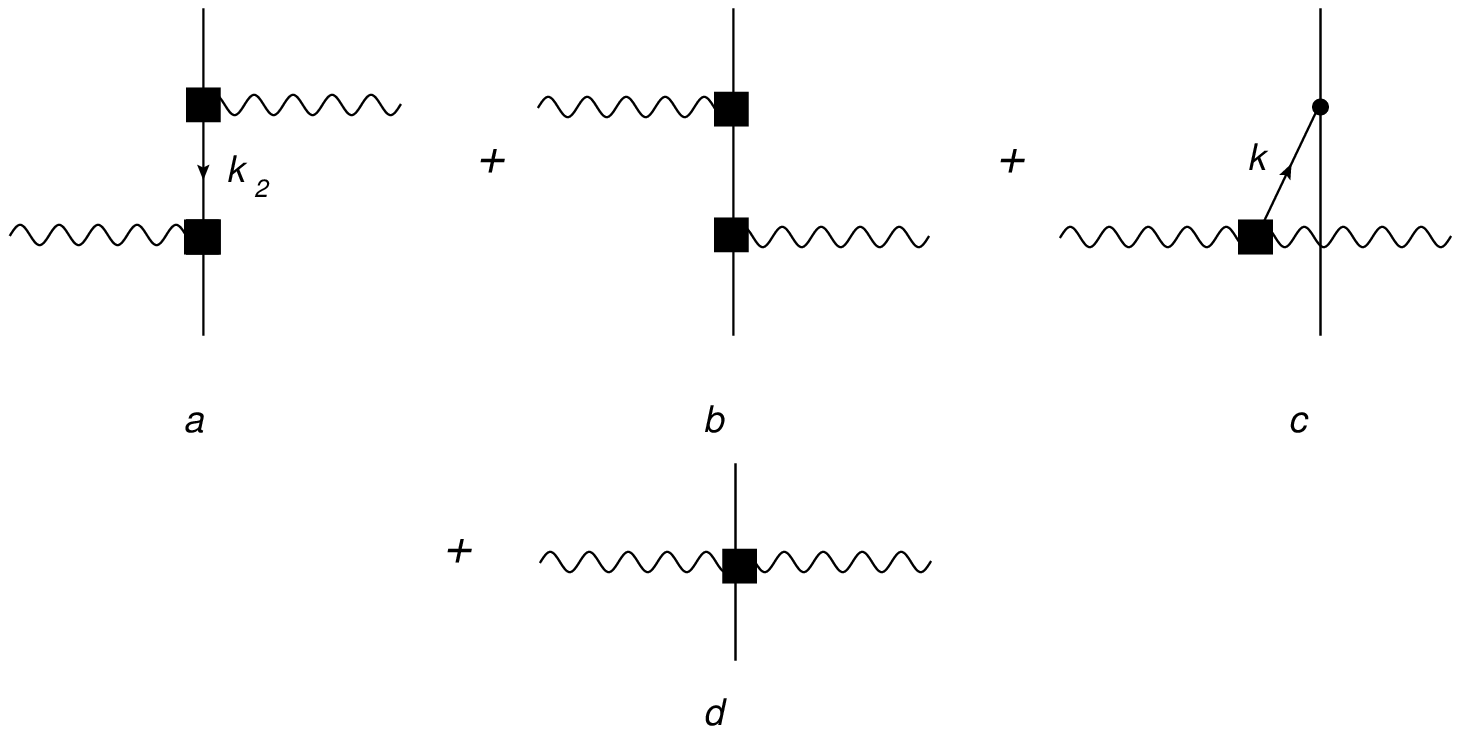,width=10cm,height=6cm}\\
Fig.6: The effective RPPR vertex
\end{center}
The sum of these four structures (a) + (b) + (c) + (d) is gauge invariant, 
i.e it satisfies the Ward identity in $k_1$, 
provided the gluon with momentum $k_3$ is on shell and we multiply with a physical polarization vector. 
Decomposing the three different color tensors in the two independent ones, making use of the Jacobi identity on the non planar contribution, we can verify that the corresponding combination of each planar structure with the non planar one separately satisfies the Ward identity.
Let us define the color stripped contribution corresponding to Fig.6a as
\bea
\label{planar-a}
I_a^{\alpha \gamma}(k_1,k_3) \hspace{-0.5cm}&{}&=  g^{\alpha\gamma} q_{2\perp}q'_{2\perp} - (q_{2\perp})^{\alpha} (q'_{2\perp})^{\gamma}+
 \\
&{}& \hspace{-1.7cm}\left[ \gamma^{\beta \alpha \mu'}( -k_2,q'_2) (q'_{2\perp}) _{\mu'}- \frac{q{'_2}^-}{2} \frac{{q'}_{2\perp}^2}{k_1^+}    (n^+)^{\alpha} (n^+)^{\beta} \right] \frac{1}{k_2^2} \left[ (q_{2\perp}) _{\mu} 
 \gamma^{\mu \beta  \gamma}(q_2,-k_3) - \frac{(q_2)^+}{2} \frac{q_{2\perp}^2}{k_3^-}    (n^-)^{\beta} (n^-)^{\gamma}  \right].
\nonumber
\eea 
Here  the first line contains the contribution of the quartic coupling, and  in the second line
we have put  $k_2=k_1-q'_2=-k_3-q_2$.
The nonplanar contribution corresponding to Fig.6c can be written as
\bea
\label{nonplanar}
I_c^{\alpha \gamma}(k_1,k_3) \hspace{-0.5cm}&{} &=  q_{2\perp}^\alpha (q'_{2\perp})^\gamma - (q_{2\perp})^{\gamma} (q'_{2\perp})^{\alpha}+
\\
&{}& \hspace{-1.7cm} \left(   (q_2)_{\perp \mu} \gamma^{\mu \beta \mu'}(q_2,q'_2) (q'_2)_{\perp \mu'} \right) 
\frac{1}{k^2} 
\gamma^{\alpha \beta \gamma}(k_1,-k_3)\,.
\nonumber
\eea 
Again, the first line contains the quartic coupling term, and in the second line we have used  $k=q_2-q'_2=-k_1-k_3$.

In terms of these quantities the Ward identity we want to check is given by
\be
k_{1 \alpha} \left(I_a^{\alpha \gamma} - I_c^{\alpha\gamma} \right) \epsilon_{3\gamma} =0,
\label{WardIdRPPR}
\ee
where $\epsilon_{3 \gamma} k_3^\gamma=0$ and $k_3^2=0$.
In order to show it one can use the following identities:
\bea
\!\!\!\!\!
&{}&k_{1\alpha} \left[ \gamma(-k_2, q'_2)^{\beta \alpha \mu'}(q'_{2\perp}) _{\mu'}- \frac{{q'_2}^-}{2} \frac{{q'}_{2\perp}^2}{k_1^+}   
(n^+)^{\alpha} (n^+)^{\beta} \right]
=-k_2^\beta q'_{2\perp} k_{2\perp}+(q'_{2\perp})^\beta k_2^2\,, \nonumber\\
&{}&k_{2\beta} \left[ (q_{2\perp}) _{\mu}  \gamma(q_2,-k_3)^{\mu \beta \gamma} - \frac{(q_2)^+}{2} \frac{q_{2\perp}^2}{k_3^-}    (n^-)^{\beta} (n^-)^{\gamma}  \right]
= k_3^\gamma q_{2\perp} k_{3\perp}-(q_{2\perp})^\gamma k_3^2\,, \nonumber\\
&{}&k_{1\alpha}\left[ g^{\alpha\gamma} (2k_3\!+\!k)^\beta\!-\!g^{\alpha \beta} (k_3\!+\!2k)^\gamma \!-\!g^{\beta\gamma} (k_3\!-\!k)^\alpha \right]
=k^\beta(k\!+\!k_3)^\gamma\!-\!(k\!+\!k_3)^\beta k_3^\gamma\!+\!(k_3^2\!-\!k^2)g^{\beta\gamma},
\nonumber
\eea
where in the first two lines (planar structure) one uses $k_2=k_1-q'_2$, and in the last line (non planar structure) $k=-k_1-k_3$.
Then one obtains
\bea
k_{1\alpha} \epsilon_{3\gamma} I_a^{\alpha \gamma}&=&\epsilon_{3\gamma}\left[ 
q'_{2\perp} k_{3\perp} (q_{2\perp})^\gamma-q_{2\perp} k_{3\perp} (q'_{2\perp})^\gamma
+ \frac{1}{2}q_{2\perp} q'_{2\perp} \left[q_2^+(n^-)^\gamma+{q'_2}^- (n^+)^\gamma\right]
\right]\,,\nonumber\\
k_{1\alpha} \epsilon_{3\gamma} I_c^{\alpha \gamma}&=&\epsilon_{3\gamma}\Bigl[
(q_2\!-\!q'_2)_\perp q_{2\perp} (q'_{2\perp})^\gamma\!-\!(q_2\!-\!q'_2)_\perp q'_{2\perp} (q_{2\perp})^\gamma+
\frac{1}{2}q_{2\perp} q'_{2\perp} \left[q_2^+(n^-)^\gamma+{q'_2}^- (n^+)^\gamma\right]\nonumber\\
&{}&\,\,\,\,\,\,\,\,\,+q_{2\perp} k_{1\perp} (q'_{2\perp})^\gamma-q'_{2\perp} k_{1\perp} (q_{2\perp})^\gamma
\Bigr],
\eea
which leads to \eqref{WardIdRPPR}.

Similarly one can write the color stripped contribution of Fig.6b 
\bea
\label{planar-b}
I_b^{\alpha \gamma}(k_1,k_3) \hspace{-0.5cm}&{}&= (q'_{2\perp})^{\alpha} (q_{2\perp})^{\gamma} -g^{\alpha\gamma} q_{2\perp}q'_{2\perp}  +
 \\
&{}& \hspace{-1.7cm}
\left[ (q_{2\perp}) _{\mu}  \gamma^{\mu \alpha \beta}(q_2,k_2) - \frac{(q_2)^+}{2} \frac{q_{2\perp}^2}{k_1^-}    (n^-)^{\alpha} (n^-)^{\beta}  \right]
  \frac{1}{k_2^2} 
\left[ \gamma^{\gamma\beta \mu'}(k_3,q'_2) (q'_{2\perp}) _{\mu'}-+\frac{{q'_2}^-}{2} \frac{{q'}_{2\perp}^2}{k_3^+}    (n^+)^{\beta} (n^+)^{\gamma} \right],
\nonumber
\eea
where $k_2=k_1+q_2=q'_2-k_3$. Combined with the nonplanar contribution Fig.6c, 
one easily verifies the Ward identity.

In the context of this paper, in which we study the NLO corrections to the BKP equation for the Odderon channel 
(and for the large-$N_c$ limit of the multi-reggeon channel), 
the nonplanar structure of Fig.6c does not contribute, and the RPPR vertex reduces to  $I_a^{\alpha \gamma}(k_1,k_3) +I_b^{\alpha \gamma}(k_1,k_3)$, 
written in eqs. \eqref{planar-a} and \eqref{planar-b}. Now the Ward identities follow from 
adding Figs.6a and b. 

\subsection{Putting things together}

Let us now combine these building blocks and obtain the numerator: $N$=RRPR $\otimes$ RPRR + quartic coupling. The prefactors are given in (\ref{prefactor}). The numerator is
proportional to $q_{1\perp}$, $q'_{1\perp}$,  $q_{2\perp}$, $q'_{2\perp}$,$q_{3\perp}$, $q'_{3\perp}$. The dependence upon 
$q_{1\perp}$, $q'_{1\perp}$ is through $q_{1\perp} \cdot q'_{1\perp}$ or $q_{1\perp} \times q'_{1\perp}$, similarly for $q_{3\perp}$ and $q'_{3\perp}$. It  is therefore convenient 
to organize the numerator in the following four groups: \\
1) terms proportional to $(q_{1\perp} \times q'_{1\perp}) (q_{3\perp} \times q'_{3\perp}$) are:
\bea
q_1^+ q_3^+ {q'_1}^- {q'_3}^- q_{2\perp} q'_{2\perp}\nonumber \\
 - q_1^+ {q'_1}^- q'_{2\perp} k_{2\perp} q_{2\perp} k_{3\perp}
+  q_1^+ {q'_3}^- q'_{2\perp} k_{1\perp} q_{2\perp} k_{3\perp}
+ q_3^+ {q'_3}^- q_{2\perp} k_{2\perp} q'_{2\perp} k_{1\perp} \nonumber\\
-(q_2^+ {q'_2}^-  + q_3^+ {q'_1}^-)  (q_{2\perp}  \times k_{3\perp} )
(q'_{2\perp}  \times k_{1\perp} ) .
\label{num1}
\eea  
Note that the sum of all terms without longitudinal momenta cancel.\\
2) terms proportional to $(q_{1\perp} \cdot q'_{1\perp}) (q_{3\perp} \times q'_{3\perp})$ are:
\bea
q_1^+  {q'_2}^- \left( q_3^+ {q'_3}^- (q_{2\perp}  \times  q'_{2\perp} ) - 
q'_{2\perp} k_{2\perp}  (q_{2\perp}  \times k_{3\perp})  -
k_{3\perp} ^2 (q'_{2\perp} \times k_{2\perp} ) -
k_{2\perp} ^2 (q'_{2\perp} \times k_{3\perp} ) \right) \nonumber \\ 
+  \frac{1}{2}  \left ( \Big[ (q_3^+ -q_2^+) {q'_1}^- - q_1^+ {q'_3}^-\Big]  (
k_1^2 - k_2^2)  +
(q_3^+ - q_2^+ ) {q'_2}^-  {q'_{2\perp} }^2 \right) (q_{2\perp}  \times k_{3\perp}).
\label{num2}
\eea
3) terms proportional to $(q_{1\perp} \times q'_{1\perp}) (q_{3\perp} \cdot q'_{3\perp}$) are:
\bea
q_2^+  {q'_3}^-\left( q_1^+ {q'_1}^-  (q_{2\perp}  \times  q'_{2\perp} ) - q_{2\perp}k_{2\perp} (q'_{2\perp}  \times k_{1\perp}) + k_{1\perp}^2  (q_{2\perp}  \times k_{2\perp})
- k_{2\perp}^2  (q_{2\perp}  \times k_{1\perp})
\right) \nonumber \\ 
 +   \frac{1}{2} 
\left ( \Big[ {q_3}^+({q'_1}^- -{q'_2}^-) - q_1^+ {q'_3}^-\Big]  (k_{3\perp}^2 - k_{2\perp}^2)
+ q_2^+ ({q'_1}^- - {q'_2}^- ) q_{2\perp}^2  \right)(q'_{2\perp}  \times k_{1\perp}).
\label{num3}
\eea
4) terms proportional to $(q_{1\perp} \cdot q'_{1\perp}) (q_{3\perp} \cdot q'_{3\perp})$ are:
\bea
 \left( \frac{1}{2}(q_3^+ {q'_1}^- + q_1^+ {q'_3}^-) (q_1^+ {q'_3}^- + k_{2\perp}^2) 
- q_1^+ q_2^+ {q'_2}^- {q'_3}^-\right)q_{2\perp}q'_{2\perp} \nonumber \\
-\frac{1}{2}(q_3^+ {q'_1}^- + q_1^+ {q'_3}^-) 
(k_{1\perp}^2 - k_{2\perp}^2)(k_{3\perp}^2 - k_{2\perp}^2) \nonumber\\
-\frac{1}{2} q_2^+ {q'_1}^- q_{2\perp}^2 (k_{1\perp}^2 - k_{2\perp}^2) \hspace{3cm}\nonumber\\
-\frac{1}{2} q_3^+ {q'_2}^- {q'_{2\perp}}^2(k_{3\perp}^2 - k_{2\perp}^2)\hspace{3cm}\nonumber \\
-\frac{1}{2} q_2^+ {q'_2}^- q_{2\perp}^2 {q'_{2\perp}}^2.\hspace{5cm}
\label{num4}
\eea

\subsection{Longitudinal integrals} 

Let us now consider the longitudinal integrations over the variables ${q'_1}^-$ and ${q'_3}^-$, keeping  ${q_1}^+$ and ${q_3}^+$ fixed. 
In the numerator we find, from (\ref{num1}) - (\ref{num4}), two groups of terms. 
The first one terms contains two powers, ${q'_2}^-{q'_3}^-$, ${q'_1}^-{q'_3}^-$, or $({q'_3}^-)^2$, the second one single powers ${q'_1}^-$, ${q'_2}^-$, or ${q'_3}^-$.
In (\ref{num1}) a third possible structure containing terms without any factors ${q'_i}^-$ could have appeared: but, as we have already mentioned, all contributions add up to zero.   Combining with the 
prefactors in (\ref{prefactor}), all these terms have in the denominator (at least) two powers of ${q'_1}^-$ and ${q'_3}^-$, and thus lead to UV-convergent integrals. It is important to 
note that the only 'dangerous combination' ${q'_1}^-{q'_2}^-$ does not appear. 
This implies that the diagrams in Fig.5 alone are convergent, i.e. we do not need 
to sum over the permutations of the reggeized gluons. 

Now let us take a closer look at the longitudinal integrals. Together with the 
denominators in (\ref{prefactor}) we are lead to compute the following integrals (using the principal value prescription, 
as explained in section 3.1) :
\bea
\label{I1}
I_1=&{\cal P}\int d{q'_1}^- d{q'_3}^- \frac{1}{D_1 D_2 D_3} \frac{1}{{q'_1}^-}\\
\label{I2}
I_2=&{\cal P}\int d{q'_1}^- d{q'_3}^- \frac{1}{D_1 D_2 D_3} \frac{1}{(q'_1+q'_3)^-}\\
\label{I3}
I_3=&{\cal P}\int d{q'_1}^- d{q'_3}^- \frac{1}{D_1 D_2 D_3} \frac{{q'_3}^-}{{q'_1}^-(q'_1+q'_3)^-}\\
\label{I4}
I_4=&{\cal P}\int d{q'_1}^- d{q'_3}^- \frac{1}{D_1 D_2 D_3} \frac{1}{(q'_1+q'_3)^-{q'_3}^-}\\
\label{I5}
I_5=&{\cal P}\int d{q'_1}^- d{q'_3}^- \frac{1}{D_1 D_2 D_3} \frac{1}{{q'_1}^-{q'_3}^-}\\
\label{I6}
I_6=&{\cal P}\int d{q'_1}^- d{q'_3}^- \frac{1}{D_1 D_2 D_3} \frac{1}{{q'_1}^-(q'_1+q'_3)^-}
\eea 
with
\bea
D_1 = k_1^2 + i \epsilon = &{q_1}^+ (-q'_1)^- + k_{1\perp}^2 + i \epsilon \\
D_2=k_2^2 +i \epsilon= &{q_1}^+ {q'_3}^- + k_{2\perp}^2 + i \epsilon \\
D_3=k_3^2+i \epsilon= &{q_3}^+ (-q'_3)^- + k_{3\perp}^2 + i \epsilon .
\eea

In each case, we have to consider all possible signs of  ${q_1}^+$ and ${q_3}^+$.
Let us denote the four regions ($q_1^+>0,\, q_3^+>0$), ($q_1^+<0,\, q_3^+<0$), ($q_1^+>0,\, q_3^+<0$), and ($q_1^+<0,\, q_3^+>0$) by $>>$, $<<$, $><$, $<>$, resp.
As we have already said before, each ${q'_1}^-$-integral is ultraviolet convergent. The integrations are done by closing contours in the upper or lower half plane.
We list the results: 
\be
I_1={\cal P}\int d{q'_1}^- d{q'_3}^- \frac{1}{D_1 D_2 D_3} \frac{1}{{q'_1}^-} =\left\{ \begin{array}
{ll} 2(\pi i)^2 \frac{1}{k_{1\perp}^2D_{23}} & >> \, \;or\; <<\\
0&>< \;or\;<>
\end{array} \right. ,
\ee  
\be
I_2={\cal P}\int d{q'_1}^- d{q'_3}^- \frac{1}{D_1 D_2 D_3} \frac{1}{(q'_1+q'_3)^-} =\left\{ \begin{array}
{ll}2(\pi i)^2 \frac{q_3^+}{D_{23} D_{13}} & >> \;or\; <<\\
0&><\;or\;<>
\end{array} \right., 
\ee  

\be
I_3={\cal P}\int d{q'_1}^- d{q'_3}^- \frac{1}{D_1 D_2 D_3} \frac{{q'_3}^-}{{q'_1}^-(q'_1+q'_3)^-} =\left\{ \begin{array}
{ll} 2(\pi i)^2 \frac{q_1^+ k_{3\perp}^2}{k_{1\perp}^2 D_{13} D_{23} } & >> \;or\; <<\\
0&><\;or\;<>
\end{array} \right. ,
\ee 
\be
I_4={\cal P}\int d{q'_1}^- d{q'_3}^- \frac{1}{D_1 D_2 D_3} \frac{1}{(q'_1+q'_3)^-{q'_3}^-} =\left\{ \begin{array}
{ll} 2(\pi i)^2 \frac{(q_3^+)^2}{k_{3\perp}^2 D_{23} D_{13}}-(\pi i)^2\frac{1}{k_{1\perp}^2 k_{2\perp}^2k_{3\perp}^2} & >> \;or\; <<\\
-(\pi i)^2\frac{1}{k_{1\perp}^2 k_{2\perp}^2k_{3\perp}^2}&><\;or\;<>
\end{array} \right. .
\ee   
\be
I_5={\cal P}\int d{q'_1}^- d{q'_3}^- \frac{1}{D_1 D_2 D_3} \frac{1}{{q'_1}^-{q'_3}^-} =\left\{ \begin{array}
{ll} 2(\pi i)^2 \frac{q_3^+}{k_{1\perp}^2 k_{3\perp}^2 D_{23}}-(\pi i)^2\frac{1}{k_{1\perp}^2 k_{2\perp}^2k_{3\perp}^2} 
& >>\, \;or\; <<\\
-(\pi i)^2\frac{1}{k_{1\perp}^2 k_{2\perp}^2k_{3\perp}^2} &><\;or\;<>
\end{array} \right. ,
\ee  
and 
\be
I_6={\cal P}\int d{q'_1}^- d{q'_3}^- \frac{1}{D_1 D_2 D_3} \frac{1}{{q'_1}^-(q'_1+q'_3)^-} =\left\{ \begin{array}
{ll} 2(\pi i)^2 \frac{q_1^+ q_3^+}{k_{1\perp}^2 D_{13} D_{23}}-(\pi i)^2\frac{1}{k_{1\perp}^2 k_{2\perp}^2k_{3\perp}^2} 
& >>\;or\; <<\\
-(\pi i)^2\frac{1}{k_{1\perp}^2 k_{2\perp}^2k_{3\perp}^2}  &><\;or\;<>
\end{array} \right. .
\ee  
with 
\be
D_{13}= q_3^+ k_{1\perp}^2 + q_1^+ k_{3\perp}^2
\ee
and
\be
D_{23}= q_3^+ k_{2\perp}^2 + q_1^+ k_{3\perp}^2.
\ee
In all cases, the upper lines on the rhs correspond to the case of equal sign: $q_1^+>0, q_3^+>0$ or  $q_1^+<0, q_3^+<0$, the lower lines to opposite signs: $q_1^+>0, q_3^+<0$ or $q_1^+<0, q_3^+>0$.

Looking at these results we make a few important observations. First, one easily verifies that in all $I_i$, for the case of equal signs,  the first term on the rhs could have been obtained by ignoring the poles from $1/(q'_1)^-$, $1/(q'_2)^-$ and $1/(q'_3)^-$, closing the contours of the  $(q'_1)^-$ and $(q'_3)^-$ integrations in the upper half planes and picking the poles in  $(q'_1)^-$  and $(q'_3)^-$ arising from the denominators $D_1$ and $D_3$, resp., and finally multiplying by a factor $1/2$. Second, the integrals $I_4$, $I_5$, and $I_6$  contain, on the rhs, terms $ \sim 1/k_{1\perp}^2 k_{2\perp}^2 k_{3\perp}^2$ 
which appear both for equal and mixed signs of the $q_i^+$. Combining with the $q_i^+$ factors contained in the prefactors (\ref{prefactor1}), their dependence upon 
$q_i^+$ is always of the simple form $1/q_i^+q_j^+$ ($i \neq j$): when integrating over 
positive and negative values of $q_i^+$ and $q_j^+$, these terms cancel. As result, we are 
left with only the first terms proportional to $1/D_{23}$ or $1/D_{13} D_{23}$, belonging to the equal sign regions.
For the integrals $I_2$, $I_3$, $I_4$, and $I_6$ we encounter the denominators $D_{23}$ and $D_{13}$: the first one results from 
inserting the ${q'_3}^-$ pole from  $D_3$ into the gluon propagator $1/D_2$, the second one from inserting 
the ${q'_3}^-$ pole (from  $D_3$) and the ${q'_1}^-$ pole (from  $D_1$) into the 
denominator $1/(q'_1+q_3)^-= - 1/{q'_2}^-$  (which had its origin in our use of the Ward identities).  

We summarize these findings as follows:\\
(i) only the equal sign regions, $<<$ and $>>$, contribute,\\
(ii) apart from the overall factor $1/2$, the results are the same as obtained from taking residues of $D_1$ and $D_3$, i.e. by putting  the lines $k_1$ and $k_3$ on-shell.
 
With these findings we can come back to the representation of the $3\rightarrow 3 $
kernel given by Fig. 4. Since the only  diagram having simultaneously poles at $D_1=0$ and $D_3=0$ is 
the first diagram on the rhs of  Fig. 4, only this diagram must be considered.  We are thus left with the diagram illustrated in 
Fig.7 with the product  
$RPR \otimes RPPR \otimes RPR$, with the s-channel gluons $k_1$ and $k_3$ being 
on-shell.  
 \begin{center}
\epsfig{file=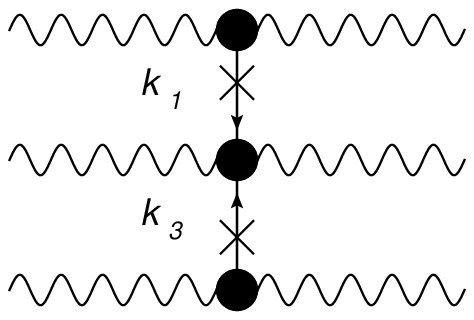,width=5cm,height=3cm}\\
Fig.7: Final form of the $3 \to 3$ kernel: product of the effective RPR, RPPR, RPR vertices
in the notation of Fig.3. The s-channel gluons with momenta $k_1$ and $k_3$ are on-shell.
\end{center}
Due to gauge invariance of the on-shell vertices we can use any gauge for the propagators of the gluons $k_1$ and $k_3$.

\end{document}